\documentclass{emulateapj}

\usepackage{natbib}
\usepackage{epsfig,color}
\usepackage{amssymb,amsmath}
\usepackage{multirow}

\shorttitle{Finding lenses near lenses}
\shortauthors{Newton et~al.}

\def\hon{$H_1$}
\def\htw{$H_2$}
\def\hth{$H_3$}
\def\hfo{$H_4$}
\def\hro{H_r^{opt}}
\def\hrr{H_r^{real}}
\def\thetaE{\theta_{\rm E}}
\def\zd{z_d}
\def\zs{z_s}

\def\Ds{D_s}
\def\Dds{D_{ds}}

\def\sigstar{\sigma_*}
\def\sigSIE{\sigma_{\rm SIE}}
\def\mf{m_{\rm F814W}}
\def\S{Section}

\begin{document}

\title{Enhanced lensing rate by clustering of massive galaxies: \\
newly discovered systems in the SLACS fields}

\author{Elisabeth R. Newton\altaffilmark{1}}
\author{Philip J. Marshall\altaffilmark{1}}
\author{Tommaso Treu\altaffilmark{1,2}}

\altaffiltext{1}{Department of Physics, University of California,
Santa Barbara, CA 93106-9530}
\altaffiltext{2}{Alfred P. Sloan Research Fellow; Packard Fellow}

% ----------------------------------------------------------------------------

\begin{abstract}

Galaxy-scale strong gravitational lens systems are useful for a
variety of astrophysical applications. However, their use is limited
by the relatively small samples of lenses known to date. It is thus
important to develop efficient ways to discover new systems both in
present and forthcoming datasets. For future large high-resolution
imaging surveys we anticipate an ever-growing need for efficiency and
for independence from spectroscopic data. In this paper, we exploit
the clustering of massive galaxies to perform a high efficiency
imaging search for gravitational lenses.  Our dataset comprises 44
fields imaged by the Hubble Space Telescope (HST) Advanced Camera for
Surveys (ACS), each of which is centered on a lens discovered by the
Strong Lens ACS Survey (SLACS). We compare four different search
methods: 1) automated detection with the HST Archive Galaxy-scale
Gravitational Lens Survey (HAGGLeS) robot, 2) examining cutout images
of bright galaxies (BGs) after subtraction of a smooth galaxy light
distribution, 3) examining the unsubtracted BG cutouts, and 4)
performing a full-frame visual inspection of the ACS images.  We
compute purity and completeness and consider investigator time for the
four algorithms, using the main SLACS lenses as a testbed. The first
and second algorithms perform the best. We present the four new lens
systems discovered during this comprehensive search, as well as one
other likely candidate. For each new lens we use the fundamental plane
to estimate the lens velocity dispersion and predict, from the
resulting lens geometry, the redshifts of the lensed sources. Two of
these new systems are found in galaxy clusters, which include the
SLACS lenses in the two respective fields. Overall we find that the
enhanced lens abundance ($30^{+24}_{-8}$ lenses/degree$^2$) is higher
than expected for random fields ($12^{+4}_{-2}$ lenses/degree$^2$ for
the COSMOS survey). Additionally, we find that the gravitational
lenses we detect are qualitatively different from those in the parent
SLACS sample: this imaging survey is largely probing higher-redshift,
and lower-mass, early-type galaxies.

\end{abstract}


% ----------------------------------------------------------------------------
 
\section{Introduction}\label{sec:intro}

Strong gravitational lensing -- when the potential of a massive
foreground object causes the formation of multiple images of a
background source -- is a powerful tool for cosmological and
astrophysical research.  Applications include measuring the mass
distributions of dark and luminous matter, and measuring cosmological
parameters via the lens geometry or abundance \citep[see e.g.\ ][for a review]{Koc++06}.  
A homogenous, well-understood sample of lenses is
required for a statistically significant study, necessitating
large-scale, systematic surveys \citep{TOG84, Bol++06,Ina++08}.  The
$\sim200$ galaxy-scale lenses known today have been found through the
numerous selection algorithms described in the next paragraph, and
include a significant number of serendipitous discoveries. Although
this has enabled substantial progress, the number is still
the limiting factor for many applications.

Most searches so far have focused on either the source or lens population,
employing a range of different strategies.  Searches targeting potential sources
have included looking for multiply-imaged radio sources -- as in the
Cosmic Lens All-Sky Survey \citep[CLASS;][]{Mye++03, Bro++03} -- and
examining known quasars \citep[e.g.][]{Mao++97,Pin++03}. The Sloan
Digital Sky Survey (SDSS) Quasar Lens Search \citep{Ogu++06, Ina++08}
selects lensed quasar candidates using two algorithms, one based on
morphology and color for small-separation images, and one based only
on color.  Of the surveys targeting potential lenses, most have
involved inspection of high resolution images. \citet{Rat++99} and
\citet{Mou++07} selected candidates by eye from HST color images of
the Extended Groth Strip (EGS) fields.  Other authors have attempted
to pre-select massive galaxies by their optical magnitude and color,
and then examine the residuals after galaxy subtraction.  This was done in the
Great Observatories Origins Deep Survey (GOODS) lens search
\citep[][]{Fas++04} and the Cosmic Evolution Survey (COSMOS) lens
search \citep[]{Fau++08}. Most recently, \citet{J08} extended this
method to lower mass galaxies, but focused on viewing large-format
arrays of unsubtracted color galaxy cutout images.  The Sloan Lens ACS
Survey \citep[]{Bol++06, Bol++08a} candidates were selected based on
spectroscopic data indicating multiple redshifts in the spectrum of
early-type (and hence massive) galaxies, then classified using this
data and visual examination of high resolution HST images (before and
after lens galaxy subtraction).

Recently, several algorithms for automated lens detection have been
developed.  Those developed by \citet{Ala06}, \citet[]{S+B07} and
\citet{K+D08} look for arcs, a common feature in both weak and strong
lensing on group and cluster scales.  \citet{Est++07} and
\citet{Bel++07} suggest mining databases for blue objects near
potential lenses, since the most common sources are faint blue
background galaxies. The RingFinder \citep[]{Cab++07} applies the same
logic to smaller image separation lenses, subtracting a rescaled red
image from a blue cutout image to dig within the lens light
distribution; it then analyzes the shapes and positions of the
remaining residuals.  Most recently the HAGGLeS automated lens
detection ``robot'' \citep[]{Mar++08} attempts to model every object
(typically selected to be Bright Red Galaxies or BRGs) as a
gravitational lens, i.e.\ as a combination of background light from
the source that is consistent with having been multiply-imaged, and
residual foreground light from the lens galaxy. The result is the
robot's quantitative prediction of how a human would have classified
the candidate.

Fortunately, we are about to enter an era when orders of magnitude
increases in the number of known lenses will be possible. In the near
future, wide-field surveys such as the proposed Joint Dark Energy
Mission (JDEM) and Euclid space missions would provide $\sim10^3 -
10^4$ square degrees of high resolution imaging data
\citep[]{Mar++08}.  Automation will be needed to examine such large
areas over manageable timescales, and to draw attention to those
systems which have a higher probability of being lenses. We also must
be prepared to proceed without the help of spectroscopic data.

We describe here just such a survey: based purely on imaging data, and with a
sufficient degree of automation. We compare the accuracy of four
methods of searching for strong gravitational lenses. Presented in
order of degree of automation, they are: (1) using the HAGGLeS robot
\citep{Mar++08}, (2) examining subtraction residuals \citep[e.g.][]{Fau++08}, (3) looking at galaxy cutouts
\citep[e.g.][]{Fas++04,J08}, and (4) performing a visual inspection of
entire fields \citep[e.g.][]{Mou++07}.

We additionally aim to make use of galaxy clustering in order to
improve the efficiency of our search, as suggested by
\citet[][]{Fas++06}.  The most likely lensing galaxies are massive
ellipticals with $0.3 < z < 1.3$ \citep[]{TOG84,Fas++04}: we may
expect that focusing on bright red galaxies (BRGs) will increase the
lens detection efficiency \citep[e.g.][]{Fas++04,Fau++08,Mar++08}. The
sources, meanwhile, are predominantly expected to be faint blue
galaxies (FBGs) at redshifts at or above 1 \citep[]{MBS05}.  Both of
these types of objects (potential lenses and potential sources) are
clustered, BRGs more strongly that FBGs.  We therefore expect that
strong gravitational lenses are also clustered; we anticipate finding
more new lenses by looking near known ones than we would otherwise.
For example, \citet[]{Fas++06} presents just such an occurrence: the
researchers discovered two additional lens candidates less than 40
arc-sec from the known lens B1608+656.

Therefore, as our dataset we use a subset of the SLACS HST/ACS fields.
The SLACS survey has discovered 70 definite galaxy-galaxy strong
lenses~\citep{Bol++08a} to date.  These definite lenses have clearly
identifiable lenses arcs or multiple images in addition to
spectroscopic data; due to the rigorous requirements for this
classification, we consider all to be confirmed lenses.  Of these, 63
are well-modeled by a singular isothermal ellipsoid (SIE) and have
lens and source redshifts along with F814W photometry for the lens
galaxy; most also have measured stellar velocity dispersions
\citep[]{Bol++08a}.  This makes the SLACS lenses the largest
homogenous sample of strong lenses to date.  Each of the fields we use
is centered on one of the known SLACS lenses (hereafter the ``main
lens''); we thus make use of gravitational lens clustering in every
field.  The majority of our fields were observed in just one filter,
meaning that we by necessity pre-select bright galaxies (BGs) rather
than BRGs. For the automated portion of our searches, we use the
HAGGLeS robot which, in addition to performing a quantitative lens
classification, creates a useful database of galaxy cutouts and
residuals.

The organization of this paper is as follows.  In
Section~\ref{sec:sample} we present the 44 ACS fields in our dataset,
and in Section~\ref{sec:methods} we outline the four search procedures
we used to find gravitational lenses in these images. To assess the
accuracy of these methods, we discuss each procedure's performance on
the main lenses in Section~\ref{sec:accuracy}.  We then present our
four new definite strong lensing systems and one likely candidate, and
investigate their physical properties and environments, in
Section~\ref{sec:results}.  After some discussion of our methods and
results in Section~\ref{sec:discussion}, we conclude in
Section~\ref{sec:conclusions}.  All magnitudes are in the AB system;
we assume a flat cosmology with $\Omega_m$=0.3, $\Omega_\Lambda$=0.7,
and H$_0$=100$h$ kms$^{-1}$Mpc$^{-1}$, where $h$=0.7 when necessary.

% ----------------------------------------------------------------------------
 
\section{Sample Selection and Object Detection}\label{sec:sample}

We use a subsample of the ACS F814W fields investigated in the Sloan
Lens ACS Survey
\citep[SLACS;][]{Bol++06,Tre++06,Koo++06,Gav++07,Bol++08a,Gav++08,Bol++08b,
Tre++08}, hereafter papers SLACS-I through VIII. The sample selection
and data reduction and analysis of the SLACS data is given in SLACS-I,
-IV and -V and will not be repeated here. We select the subset of 44
definite lenses whose ACS observations meet the following two
requirements: exposure time $>$1000 seconds through F814W, and lens
classification ``definite.''  Due to the demise of ACS, two filters
are not available across all fields; of the ACS filters used, F814W is
the most widely available in the SLACS fields. 14 fields were
additionally imaged through F555W.  We require long exposure time to
guarantee the ability to produce deep, high quality, cosmic ray-free
reduced images.

Our final sample comprises 44 uniformly-observed ACS fields, each
centered on a SLACS gravitational lens (a ``main lens'').  Each field
covers approximately 11 arcmin$^2$; thus our total survey area is
0.134 square degrees.

Most lens-finding methods involve examining bright galaxies (BGs), as
being the most likely to be lensing galaxies \citep[e.g.][]{Fas++04,
Fau++08, Mar++08}.  Here this selection is done with magnitude and
size cuts on catalogs made with the
SExtractor\footnote{http://terapix.iap.fr/} software tool; only a
third of our fields have multi-filter ACS data so in order to maintain
consistency we do not use color selection even when it is possible.
We choose an apparent F814W magnitude ($\mf$; using Kron-like elliptical
apertures) limit of 22. Other limits were tried, but 22
was found empirically to be the best balance between efficiency and
the selection of all bright galaxies: with $\mf<22$, a total of 2399
BGs are selected, but with $\mf<23$, that number jumps to 5439
without adding significantly to the actual number of morphologically
early-type galaxies selected.  The minimum size selected is
FWHM=0$\farcs$215, in order to reject most stars.  We additionally use
the SExtractor ``flags'' parameter to limit the remaining
contaminants.

% ----------------------------------------------------------------------------

\section{Search Methods}\label{sec:methods}

We use four distinct methods to search for gravitational lenses.  They
are presented in order of degree of automation, where the first has the most
pre-selection and analysis, and the last none:

\begin{itemize}
\item Procedure 1: Using output from the HAGGLeS robot
\item Procedure 2: Inspection of subtraction residuals of bright galaxies
\item Procedure 3: Inspection of cutout images of bright galaxies
\item Procedure 4: Full-frame visual inspection of the ACS images
\end{itemize}

In general we expect our searches to be sensitive to
lensing events similar to, but fainter than, the SLACS main
lenses. With the BG selection in Procedures 1--3, and the innate
limits of the human eye in Procedure 4, we are biased towards lensing
events around large, luminous galaxies. However, empirically the main
lenses are the brightest objects (other than saturated stars) in the
fields, so we expect any other lenses we find to be, by necessity,
fainter. The main lenses also favor high magnification configurations
(Einstein rings) due presumably to their spectroscopic selection; we
expect to be somewhat preferential in this regard as well, on the
grounds that these distinctive cases will be easiest to identify
visually.

Before continuing, we first consider the question ``what is the
probability that this object is a gravitational lens?''  The
classification of lens candidates has varied among the major surveys
targeting potential lenses, as has the data on which these
classifications are based.  The COSMOS lens search \citep[]{Fau++08}
sample is subdivided into ``best systems'', which are deemed to have a
greater probability of being lenses than the remaining objects.  GOODS
\citep[]{Fas++04} selected candidates and voted to choose ``top
candidates.''  \citet[]{J08} groups potential lenses into three
categories: very likely or certain lenses, possible or probable
lenses, and not-lenses.  The SLACS survey \citep[]{Bol++08a} also
groups candidates into three categories (definite lenses, probable
lenses, and inconclusive/not lenses).  We note that the classification
of definite or probable lenses is not the same across surveys, and
therefore one has to be careful to impose similar quality criteria
when comparing inferred density of lens galaxies.

\begin{table}
\caption{Lens Classification Systems}\label{tab:class}
\begin{center}\begin{tabular}{ccc}
\hline\hline
Human class $H$ & Robot class $H_r$ & Definition \\
\hline
3 & $>$2.5 & Definitely a lens \\
2 & 1.5 to 2.5 & Probably a lens \\
1 & 0.5 to 1.5 & Possibly a lens \\
0 & $<$0.5 & Definitely not a lens \\
\hline
\end{tabular}\end{center}\end{table}

In this work we follow \citet{Mar++08} and employ a 4-point subjective
classification scheme, outlined in Table~\ref{tab:class}: the classification 
parameter $H$ may range from 3 (definite lenses), through 2 (probable lenses)
and 1 (possible lenses), to $H=0$  (definitely not lenses).  Each of the 44
fields in our sample contains a confirmed  lens from the SLACS survey,
where here the grade is based on all available data, notably the  galaxy
spectrum (including anomalous high redshift emission lines) and clear
lensing-consistent residuals after lens galaxy light subtraction in all procured filters.  
We thus assign the classification parameter $H=3$ for each main lens. 

Humans (after some training) are adept at identifying gravitational
lenses by eye, using high resolution imaging data alone; we
effectively make an internal model for the lens and optimize
it. However, the amount of information available during each lens
search with which to do this may vary: in any given search we always
lack additional understanding that would aid us in identifying lenses.
When carrying out the four different search procedures, we therefore
assign each system a value $H_i$ (where $i=1,2,3$ or $4$): each of
these is our best guess as to the classification that a trained human
would have assigned the system, if they had been given only the data
presented in the $i^{\rm th}$ procedure.  (This results in four
``re-classifications'' for each main lens.)  We consider the ``true''
classification $H$ to be the value we give an object when taking into
account \emph{all data}.

% - - - - - - - - - - - - - - - - - - - - - - - - - - - - - - - - - - - - - - - - 

\subsection{Procedure 1: HAGGLeS robot}

In Procedure 1, we use the HAGGLeS robot, the automated lens detection program
developed by \citet[]{Mar++08}, to identify samples of lens candidates prior
to visual inspection.  The robot treats every object as if it were a
lens, models it, and then calculates how well the lens hypothesis works.  It
extracts and uses 6 arc-second square cutout images of each object, and then
focuses on the residuals made by subtracting off an elliptically symmetric
Moffat profile model for the putative lens galaxy light. The gravitational
potential of the lensing galaxy is assumed to be sufficiently well-described
by a singular isothermal sphere (SIS) plus external shear; this model is
fitted to the residual image. Disk-like features and bright neighboring
objects are masked before fitting.  When multi-filter data is available it is
used, and although the robot does not rely on color, with it performance
is expected to improve \citep[]{Mar++08}.  Based on the results of the
modeling process, the robot calculates and assigns a value of $H_r$ to each
object, where $H_r$ is the robot's estimate of the classification $H$ a human
would have given the system.  We summarize the human and robot classifications
in Table \ref{tab:class} \citep[after][]{Mar++08}.

The robot may have one of two characters, reflecting the prior
probability that an object is a lens.  The ``realistic'' character
robot might expect 0.1\% of objects to be lenses, and the
``optimistic'' character robot might expect 60\% to be
\citep[]{Mar++08}.  The realistic robot is approximately in accordance
with current estimates of the fraction of strongly lensed galaxies
\citep[e.g.][]{Bol++06,Mou++07,Mar++08}, while the optimistic robot
has the advantage of being most inclusive by giving higher values of
$H_r$.  These two prior probability distributions are given
in Table \ref{tab:pdf} \citep[after][]{Mar++08}. \citet{Mar++08} note
that while the realistic robot produces lens samples of high purity -- and
correspondingly little need for human inspection -- a highly
complete search requires, at present, a more optimistic robot.

\begin{table}
\caption{Prior probability distributions for robot characters}\label{tab:pdf}
\begin{center}\begin{tabular}{ccccc}
\hline\hline
Character&  Pr($H$=0) & Pr($H$=1) & Pr($H$=2) & Pr($H$=3) \\ \hline
Realistic&  0.900 &     0.080 &     0.019     & 0.001 \\
Optimistic& 0.050 &     0.100 &     0.250     & 0.600 \\
\hline
\end{tabular}\end{center}\end{table}

We ran both the optimistic and realistic robots in order to verify
this last claim, but chose to use the optimistic robot to search for
new lenses, as our sample is small and we are most interested in
finding all lenses present in our sample. When comparing the
performances of both robots, we refer to the optimistic and realistic
robot classifications as $\hro$ and $\hrr$, respectively.  The main
goal of the comparison is to gain insight into the populations of
objects selected, and thus help improve the automated part of this
procedure for future surveys.

The output from the robot includes an individualized webpage for each object,
containing the cutout, subtraction residual, masked residual, and estimation
of Einstein radius.  This page also displays the minimal source  able to
produce the observed configuration, and the predicted image-plane residual
reconstructed from this minimal source \citep[see][for more details]{Mar++08}.   
We viewed the pages of objects for which $\hro > 1.5$ (indicating optimistically 
probable and definite lenses),  and gave each
object a human classification \hon, using the same data that was available to
the robot (namely, the cutout image and subtraction residual image), and the
lensmodel outputs produced by the robot.

% - - - - - - - - - - - - - - - - - - - - - - - - - - - - - - - - - - - - - - - - 

\subsection{Procedure 2: Examining BG residuals}

Procedure 2 makes use only of the subtraction residuals produced by
the robot.  We examine the residuals of every BG initially selected in
the fields; all of the $6''$ by $6''$ subtraction residual cutouts for
a field are displayed in a grid for rapid viewing. When multi-filter
data is available, the residuals are shown in color. Looking at the 44
grids, we note all objects of interest; to each of these objects we
assign a human classification \htw, based solely on the subtraction
residual data.

% - - - - - - - - - - - - - - - - - - - - - - - - - - - - - - - - - - - - - - - - 

\subsection{Procedure 3: Examining BG cutouts}

In this method, we inspect each of the $6''$ by $6''$ cutouts of the BGs
provided by the HAGGLeS robot. We again display the cutout images in a grid
for rapid viewing, making this procedure very similar to that employed by
\citet{J08}.  However, the stretch of the cutouts' image display is fixed for
all objects at a level appropriate to the EGS-type lenses \citep{Mou++07}.
This makes this method particularly sensitive to faint lensing events, and
somewhat insensitive to lensing by very bright galaxies.   We assign objects of interest a
human classification, this time \hth, using only the galaxy cutout on display.

% - - - - - - - - - - - - - - - - - - - - - - - - - - - - - - - - - - - - - - - - 

\subsection{Procedure 4: Full-frame visual inspection}

During the full-frame visual inspection, each of the 44 fields' F814W ACS
images is viewed with ds9.\footnote{http://hea-www.harvard.edu/RD/ds9/}  We
initially set image parameters in each field so as to
be sensitive to the lensing events we expect to find: similar to, but fainter
than, the main lenses.  We do this by setting the intensity stretch such that the main lens is slightly over-saturated;
from this display we select objects of interest. For each object, we adjust the scale limits and intensity stretch
such that the potential lensing features are most apparent, then decide on a
classification. Procedure 4 differs from the previous three most notably in that all galaxies in the
field -- not just BGs -- are examined, and that their viewing parameters are
set individually. The human classification parameter assigned in this
procedure is denoted \hfo.

% - - - - - - - - - - - - - - - - - - - - - - - - - - - - - - - - - - - - - - - - 

\subsection{Summary of Results}

During each procedure, any potential gravitational lens was marked and
assigned a human classification based only on the information
available during the search procedure by which it was found. Following
the completion of all searches, all available information was
considered holistically and objects were assigned a final human
classification $H$. The classification is based primarily on our
ability to recognize a typical lens geometry in the observed
morphology (in both the cutout and subtraction residual), on colors
and surface brightnesses consistent with lensing, and on the robot's
ability to model the object.  In total we discovered four new objects
with $H = 3$, and one object with $H = 2$. These new gravitational
lenses are presented in Section~\ref{sec:results}.

% ----------------------------------------------------------------------------

\section{Method Accuracy}\label{sec:accuracy}

Before investigating the properties of our newly-discovered lenses, we
first discuss the performance of the four search procedures in terms
of both their ability to find SLACS-type lenses (low redshift, high
magnification), and their contribution to the finding of the new
gravitational lens systems.  For each method, we calculate the purity
and completeness of the samples selected, considering our five best
systems and the SLACS main lenses together as lens systems to be
recovered.  We define purity as the percent of selected objects
(i.e. having classification parameter $H_i$ above some threshold),
that actually have final human classification $H$ greater than the
same threshold.  Only the SLACS main lenses and our 5 best systems
have $H > 1.5$, and all but one of these has $H = 3$.  Similarly,
completeness is defined as the percent of the objects with final human
class $H$ greater than some threshold that were given procedure
classification $H_i$ greater than the same threshold.  For Procedure
1, any object that was not examined because $\hro < 1.5$ is considered
to have $H_1 = 0$.  Purity and completeness were calculated for each
procedure, and for both the realistic and optimistic robots. These
statistics are summarized in Table~\ref{tab:stats}; errors are of
order a few percent.

\begin{table}
\caption{Purity and completeness for each method}
\label{tab:stats}
% \begin{minipage}{\linewidth}
% \renewcommand\thefootnote{\thempfootnote}
% \begin{center}
% \begin{tabular}{cccc}
% \hline\hline
% Procedure  & $H_i$ cut     &   Purity & Completeness   \\ \hline
% 1          & \hon$>2.5$    &     1.0  &  0.67$\pm0.08$ \\
%            & \hon$>1.5$    &     1.0  &  0.82$\pm0.06$ \\
% 2          & \htw$>2.5$    &     1.0  &  0.69$\pm0.08$ \\
%            & \htw$>1.5$    &     1.0  &  0.88$\pm0.07$ \\
% 3\footnote{New candidates only} 
%            & \hth$>2.5$    &      --  &  0             \\
%            & \hth$>1.5$    &    0.75$^{+0.20}_{-0.37}$& 0.60$^{+0.25}_{-0.30}$\\
% 4          & \hfo$>2.5$    &     1.0  &  0.88$\pm0.06$ \\
%            & \hfo$>1.5$    &     1.0  &  0.92$\pm0.06$ \\
% \hline
% Rea. robot & \hrr$>2.5$    &	  0.40$\pm0.22$ &         0.08$\pm0.06$\\
% 	     & \hrr$>1.5$    &	  0.08$\pm0.02$ &         0.41$\pm0.08$\\
% Opt. robot & \hro$>2.5$    &	  0.05$\pm0.01$ &         0.29$\pm0.08$\\
% 	     & \hro$>1.5$    &	  0.03$^{+0.02}_{-0.00}$& 0.84$\pm0.07$\\
% \hline
% \end{tabular}
% \end{center}
% \end{minipage}

%Oct 14 , note: EN, I have 30 main lenses H1 > 2.5, and 5 1.5 < H1 < 2.5 and have double
%checked this by manually counting on the table in the current incarnation
%1           & \hon$>2.5$    &     100      &   67 \\  (to 69)            
%& \hth$>1.5$    &      92      &   22 \\ (to 93-- a difference in rounding I believe)
\begin{minipage}{\linewidth}
\renewcommand\thefootnote{\thempfootnote}
\begin{center}
\begin{tabular}{cccc}
\hline\hline
Procedure   & $H_i$ cut     &   Purity(\%) & Completeness(\%)   \\ \hline                 
1           & $H_1 > 2.5$    &     100      &   69 \\                        
            & $H_1 > 1.5$    &     100      &   80 \\                        
2           & $H_2 > 2.5$    &     100      &   77 \\                        
            & $H_2 > 1.5$    &     100      &   90 \\                        
3           & $H_3 > 2.5$    &     100      &    4 \\                        
            & $H_3 > 1.5$    &      93      &   22 \\ 
4           & $H_4 > 2.5$    &     100      &   81 \\                        
            & $H_4 > 1.5$    &     100      &   92 \\                        
\hline
Rea.\ robot & $\hrr > 2.5$    &	     40      &    8 \\
	    & $\hrr > 1.5$    &	      8      &   41 \\
Opt.\ robot & $\hro > 2.5$    &	      5      &   27 \\
	    & $\hro > 1.5$    &	      3      &   82 \\
\hline
\end{tabular}
\end{center}
\end{minipage}

\end{table}

Note that during each search procedure, the main SLACS lenses
themselves ae classified, resulting in values of \hon, \htw, \hth, and
$H_4$ for each.  A lens-by-lens comparison of classification values is
available in the Appendix, while data on the total numbers of main
lenses and new lens recovered by each search procedure are presented
in Table \ref{tab:methods}.  In this table we also include the amount
of time spent per field for each method, and include the performance
of the HAGGLeS robot (with no human inspection) for comparison.

\begin{table*}
\caption{Summary of results for each method}
\label{tab:methods}
% \begin{tabular}{ccccc}
% \hline\hline
% Procedure & 1 & 2 & 3 & 4 \\ \hline
% Main lenses $H_i>1.5$  & 36 & 39 & -- & 42 \\
% Main lenses \hro$>1.5$ & 37 & -- & -- & --\\
% Main lenses \hrr$>1.5$ & 20 & -- & -- & --\\
% Total new candidates   & 46 & 24 & 11 & 54\\ 
% Total Candidates $H_i=3$& 3 & 3  & 0  & 3\\
% Total Candidates $H_i=2$& 1 & 1  & 4  & 0\\
% Typical candidates & arcs            & spirals, arcs   & blue images  & spirals, arcs \\
%                    & multiple images & multiple images & companions   & companions\\
% Time per field (min) 		& 2-4 &   1-2 & 1-2 & 20-40 \\
% Time per degree$^2$ (hours) 	& 10-20 & 5-10& 5-10& 100-200 \\
% \hline
% \end{tabular}
% 
\begin{tabular}{ccccccc}
\hline\hline
Procedure               &   Rea.\ robot & Opt.\ robot &  1 &  2 &  3 &  4 \\ 
\hline
Main lenses $H_i > 1.5$   &      20       &    36       & 35 & 39 &  8 & 42 \\
Main lenses $H_i > 2.5$   &       4       &    10       & 30 & 34 &  2 & 36 \\
New lenses  $H_i > 1.5$   &       0       &     4       &  4 &  5 &  3 &  3 \\
New lenses  $H_i > 2.5$   &       0       &     3       &  3 &  3 &  0 &  3 \\
Typical candidates      &  &   & arcs            & spirals, arcs   & blue images  & spirals, arcs \\
                        &  &   & multiple images & multiple images & companions   & companions\\
Time per field (min) 	&  &   & 2-4 &   1-2 & 1-2 & 20-40 \\
Time per degree$^2$ (hours)& & & 10-20 & 5-10& 5-10& 100-200 \\
\hline\hline
\end{tabular}

\end{table*}

In the subsections that follow, we briefly discuss these results and their
implications for optimizing lens discovery in high resolution imaging surveys.
In the Appendix we show, for reference, some of the images that comprise the
input data to the first three search procedures, namely, cutout images of the
main lenses, and of the $\hrr > 2.5$ candidates.

% - - - - - - - - - - - - - - - - - - - - - - - - - - - - - - - - - - - - - - - - 

\subsection{Procedure 1}\label{subsec:proc1a}

Inspecting all objects classified by the optimistic robot as $\hro >
1.5$, we classified three out of the five new systems as $H_1 = 3$,
and one system as $H_1 = 2$.  The final system had been assigned a
robot classification parameter of $\hro < 1.5$ and so was not
examined. This failure is most likely due to the non-lensed light
remaining after the Moffat subtraction (see the Appendix for more
illustration of this).

Indeed, for the main lenses, we found that the unmasked disk features
are the most common cause of robot detection failures: we show 
the subtraction residuals and reconstructed images in the
Appendix.  The robot may fail to find a suitable model entirely 
in some extreme cases, such as when considering objects with very strong
disk components or bright companion galaxies (main lenses SDSSJ1029+0420, SDSJ1103+5322 and SDSSJ1416+5136).
We find overall that 8 of the main lenses, including the three cases just mentioned, are not detected at
$\hro > 1.5$ by the robot.  Interestingly, one of the 36 robot-detected
systems, and three of the 8 robot-rejected systems were
human-classified as $H_1 < 1.5$: given just the output from the robot, these lenses (SDSSJ1213+6708, and SDSSJ1016+3859
SDSSJ1029+0420, and SDSSJ1032+5322) would not have been identified as
lenses by a human inspector.

%Indeed, for the main lenses, we found that the unmasked disk features
%are the most common cause of robot detection failures. We show some
%subtraction residuals, and reconstructed images in
%Figure~\ref{fig:mainlenses}.  In these cases, the robot failed to find
%a suitable model entirely.  Two of these situations (main lenses
%SDSSJ1029+0420 and SDSJ1103+5322) were lens galaxies with very strong
%disk components, and the disk mask also masked the inner lensed
%images.  In the third system (main lens SDSSJ1416+5136), the masking
%of both a small, but bright, disk component and a bright companion
%galaxy masked the lensing event.  

The lens galaxy subtraction also seems to be the major cause of false
detections: ring-like and disk-like features left over from the Moffat
profile subtraction and then incompletely masked can be wrongly
interpreted by the robot as lenses.  Examples of such false positives
can be seen in the 10 objects classified by the realistic robot as
$\hrr > 1.5$ (in the Appendix).

\begin{figure*}
\centering\epsfig{file=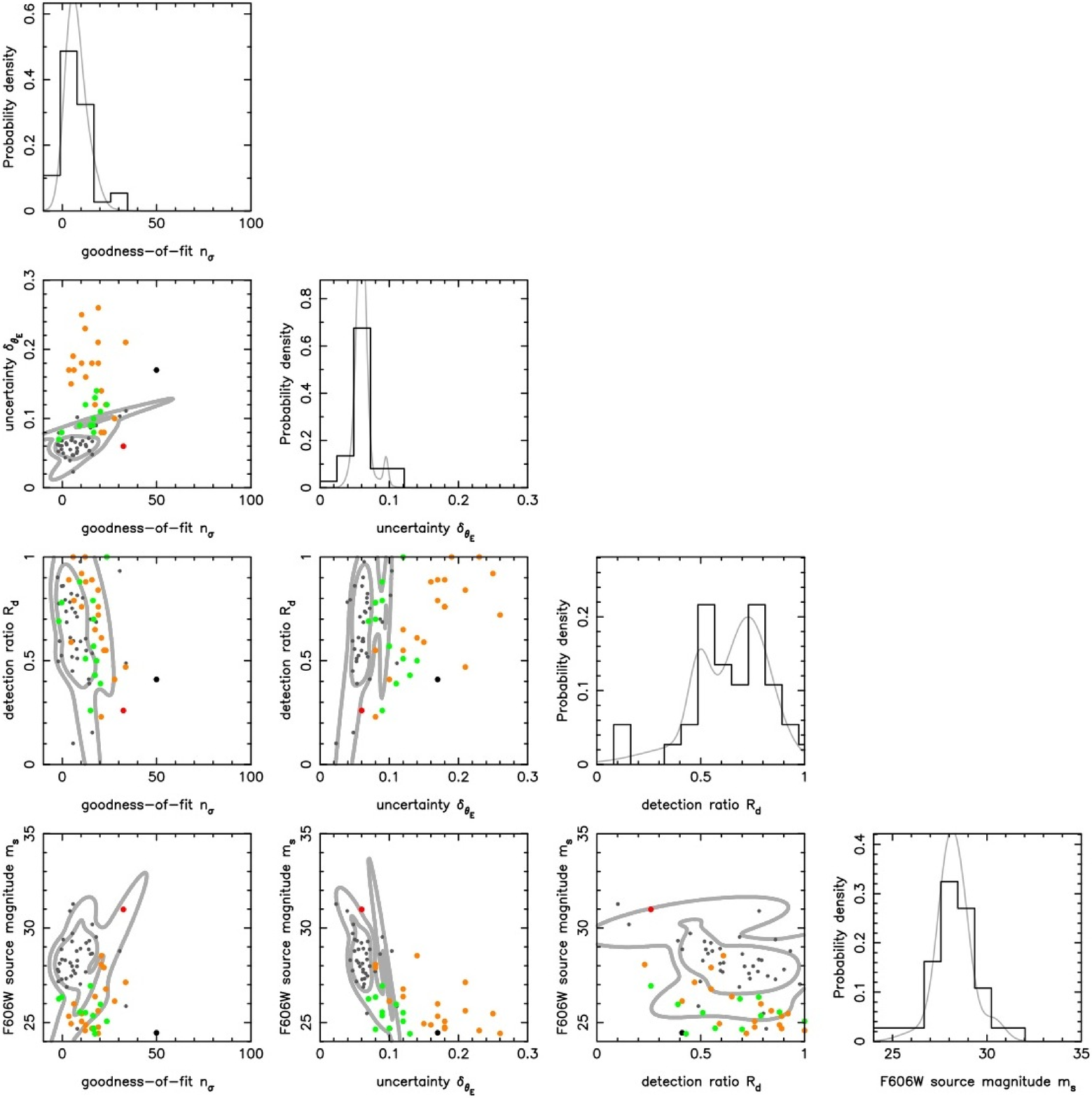,width=0.9\linewidth}
\caption{Pr(\bf{d}\vline $H$=3) derived from the robot outputs of the EGS training set (smaller points), overlain with the outputs of the SLACS main lenses (larger points).  Points correspond to objects with \hon=3 and the contours in this and the subsequent two figures are 68\% and 95\% CL.. After \citet[]{Mar++08}.}
\label{fig:priors3}
\end{figure*}

Of the 36 robot-detected main lenses, only 20 were classified by the
optimistic robot as ``definite lenses'' ($\hro > 2.5$). The reasons
for these mis-classifications are a little more subtle. The probability
density functions (PDFs) used to calculate the $H_r$ values for
this research were determined by \citet[]{Mar++08} based on a training
set of EGS non-lenses and simulated lenses.  By overlaying the robot
model output for the SLACS main lenses on these PDFs, we can gain
insight into the cause of of the robot's mis-classifications.  In
Figure~\ref{fig:priors3} we show ${\rm Pr}(\mathbf{d}|H=3)$, and
overlay the $\mathbf{d}$-values for the main lenses (larger data
points). Here we clearly see that the PDF (contours approximating the
density of smaller, training set points) is not optimized for
SLACS-type (low redshift, high magnification) lenses: to the robot,
source magnitudes are surprisingly bright and arcs are of unusual
thickness.  These differences are to be expected, but we note that the
lenses we should expect to find with the HAGGLeS robot would therefore
be more similar to the EGS lenses in terms of apparent magnitude and
geometric configuration.  In a future wide-field search, where more
SLACS-type lenses may be present, it would be prudent to retrain the
robot on a wider variety of lenses -- and perhaps on the SLACS sample
itself. %should more SLACS lenses be desired

%in particular, the source magnitudes are surprisingly (to the
%robot) bright.  Many main lenses seem also to have been misclassified
%as a result of the robot assigning large uncertainty to the Einstein
%radius, an effect which likely due to the unusual thickness (from the
%robot's point of view) of the arcs, or to the incomplete galaxy
%subtractions discussed earlier.  

\begin{figure} 
\centering\epsfig{file=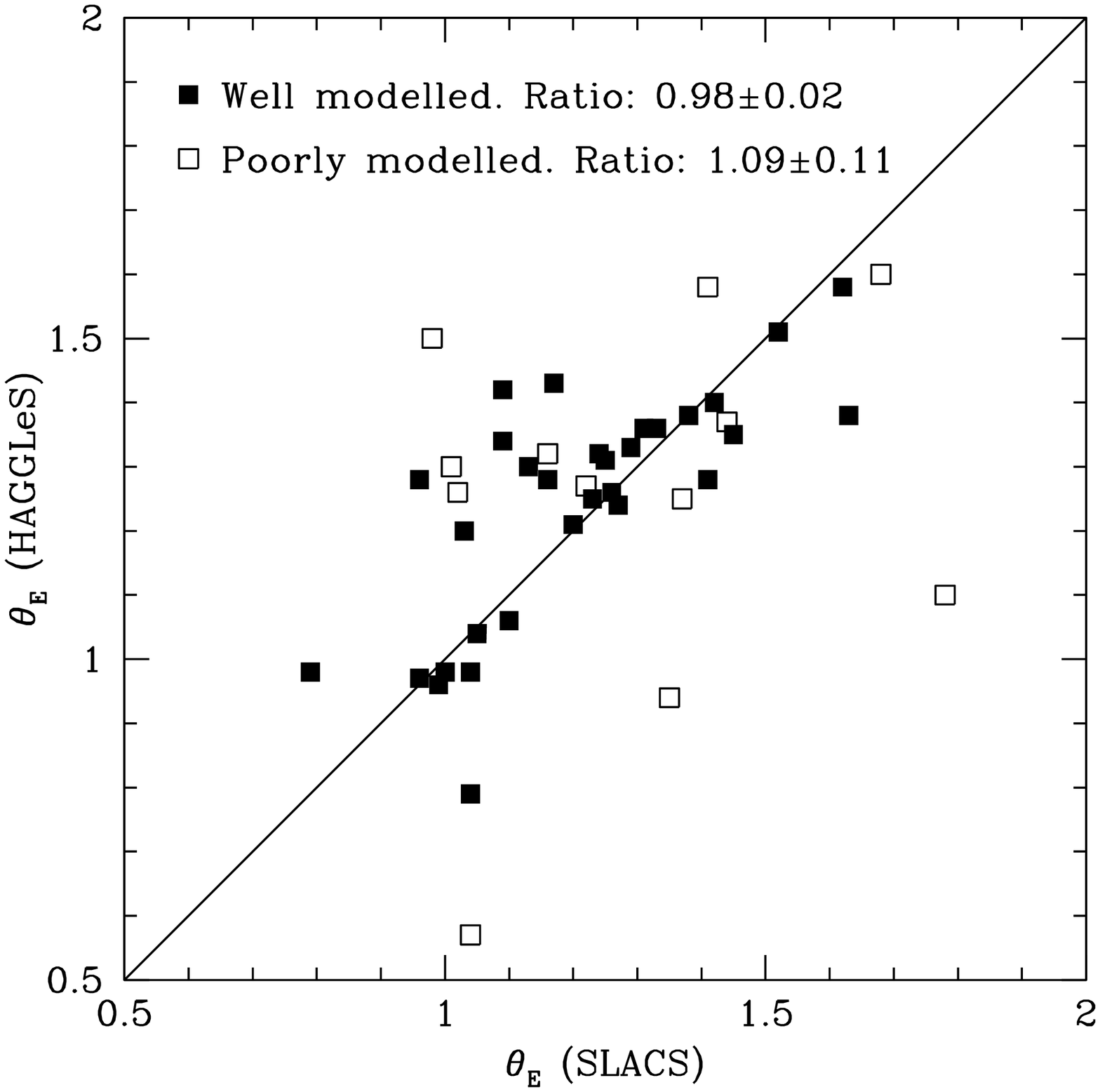,width=0.9\linewidth}
\caption{Comparison between the SLACS-measured Einstein radii (SLACS-V) and
those calculated by the HAGGLeS robot.  The systems for which robotic lens
modeling failed are represented by filled points. The one-to-one relation is
also shown to guide the eye; we find an rms scatter of 12\%.
 Errors on SLACS Einstein Radii are 2\% (SLACS-V).} \label{fig:cfrrein} 
\end{figure}

Having detected a SLACS main lens, how accurately does the HAGGLeS
robot model it?  As an illustration of the robotic models'
performance, we compare the robot-calculated Einstein radii ($\thetaE$)
to those determined in the SLACS papers (Figure \ref{fig:cfrrein}).
The 12 open data points are main lenses that were not well-modeled by
the robot; after removing these we find an rms scatter of 12\%. For
60\% of the main lenses, then, the robot finds not only a successful
lens model, but one that is in very good quantitative agreement with
that inferred during the SLACS project.

% - - - - - - - - - - - - - - - - - - - - - - - - - - - - - - - - - - - - - - - - 

\subsection{Procedure 2}

When examining the BG subtraction residual images, we classified all
four of the new $H = 3$ systems as $H_2 > 1.5$ or above; the $H = 2$
system was missed.  Einstein rings are particularly easy to identify,
resulting in a high proportion of the main lenses found.  However, we
occasionally lost some image context: some residuals are more easily
identified as being caused by lensing when viewed along side of the
lens galaxy cutout. When looking at both cutouts and subtraction
residuals, we are able to ask the question ``Is this structure a part
of the lensing galaxy or is it unique?''  However, since sources are
typically bluer than the lens galaxy, color residuals can aid us in
answering this question: we note that all but one main lens was
classified as $H_2 = 3$ when multiple filters were available.

% - - - - - - - - - - - - - - - - - - - - - - - - - - - - - - - - - - - - - - - - 

\subsection{Procedure 3}

We found that Procedure 3 was not an appropriate method to use when looking
for occurrences like the SLACS lenses.  This is predominantly an issue of image
intensity scale and dynamic range. The scales used by the robot are set for
fainter lens galaxies than those in the SLACS sample, and as such the main
lenses often appear saturated: the lensed features were completely washed
out in all but 8 cases.   We consequently classified only two objects as
$H_3 = 3$,  and this is the only procedure in which we falsely identified an
object as a lens.  This procedure was found to be most effective when lensed
images are blue against a red galaxy: five of the eight main lenses given
$H_3 > 1.5$ were in color fields.   The issue of intensity scaling was noted by
\citet{J08}, who attempted to optimize the intensity scale on an object by
object basis, and hence performed rather better in terms of ``true'' lenses
recovered (\citeauthor{J08} recovered $\sim50\%$ of the lenses identified by
\citeauthor{Fau++08}~\citeyear{Fau++08}). However, the large dynamic range in
surface brightness of the lens and source galaxies makes this a very difficult
task.

% - - - - - - - - - - - - - - - - - - - - - - - - - - - - - - - - - - - - - - - - 

\subsection{Procedure 4}

In Procedure 4 we found three out of four new $H = 3$ lenses, and only two of
the main lenses were not classified as $H_4 > 1.5$.  The greatest strength of
this method lies in the ability to vary the intensity scale of the image so as
to make apparent any lensed structures.  It is effectively the same as looking
at both a cutout and a subtraction residual: lensing can be viewed in the
context of the host galaxy.  Its weaknesses, however, are in the time
required  and the quantity of data presented: it is difficult to conduct a
thorough, mistake-free search when every object must be examined and the time
required is $\sim$30 minutes per field.

% - - - - - - - - - - - - - - - - - - - - - - - - - - - - - - - - - - - - - - - - 

\subsection{Considerations and comparisons}

We find that procedures one and two are nearly identical in their
ability to find new lenses, while the latter performed marginally
better on the main lenses since all the objects were examined.
Procedure 3 we find unsuited to the task at hand; looking at the robot
output (Procedure 1) or subtraction residuals (Procedure 2) largely
circumvents the dynamic range problems of image display, and produces
superior results.  Procedure 4 is the most inclusive of the four
procedures as all galaxies (not just bright galaxies) are examined; it
is also the only procedure in which we would have a chance of finding
a ``dark lens.''  However, we found two additional lensing systems
when we pre-selected bright galaxies and our search took significantly
less time.  We hence find that looking only at the BG subsample does
not decrease the completeness of a survey, and could in fact improve
it due to a lower error-rate.

Taking into consideration efficiency, completeness, and purity, we
recommend the use of Procedure 1 and Procedure 2. There will be cases,
such as with eight of the main lenses and one of our new lenses, where
a lens will be missed by the current robot and thus by the human
following Procedure 1.  There are also certain lenses, particularly
very unusual ones, that the current robot will miss due to its
inability to model it.  For example, the naked cusp configuration that
often occurs with edge-on spirals produces three images blended into
an arc but no counter-image; without a counter image the current robot
will classify the system as a class 0.  Additionally, when the
environment plays a strong role -- in over-dense environments for
instance -- the simple SIS+external shear model used now may not be
sufficient. It therefore may be preferential to use a combination of
procedures 1 and 2, inspecting \emph{all} objects modeled by the
robot regardless of the modeling outcome.

% ----------------------------------------------------------------------------

\section{Results}\label{sec:results}

We have discovered four new definite gravitational lenses and one
promising lens candidate. These five systems were each found in at least one
of the four separate search procedures. Having taken into
consideration all available data--including spectroscopy for one
system--we assigned four objects true human classification $H = 3$
(our four new lenses), and only one $H = 2$ (our best candidate).  The
five systems were also classified during each of the procedures.
Classifications for potential lenses are referred to as \hon, \htw,
\hth, or $H_4$ according to procedure number; these values may differ
from $H$. We present the systems along with their classifications in
Table \ref{tab:candclass}. For simplicity, each lens is given a short
name that will be used in the remainder of this paper.

\begin{table*}
\caption{Classification of new candidates by method}
\label{tab:candclass}
\begin{minipage}{\linewidth}
\renewcommand\thefootnote{\thempfootnote}
\begin{center}
\begin{tabular}{ccccccccc}
\hline\hline
Short name& Object name               &   H& \hon&    \htw& \hth& \hfo& $\hrr$& $\hro$ \\ \hline
Danny&      HST\ J073736.40+321540.3  &   3&    3&       3&    2&    3&  0.9&  1.9 \\
Frenchie&   HST\ J114331.46$-$014508.0&   3&    0$^a$ &  2&    2&    0&  0.1&  1.1 \\
Kenickie&   HST\ J121346.57+670833.3  &   3&    3&       3&    2&    3&  0.7&  2.9 \\
Sandy&      HST\ J143001.28+410440.8  &   3&    3&       3&    1&    3&  0.2&  2.9 \\
Rizzo&      HST\ J110307.14+532042.6  &   2&    2&       2&    1&    0&  0.5&  2.7 \\
\hline
\multicolumn{9}{l}{\scriptsize\raggedright $^a$ Frenchie was not examined
during procedure 1 because $\hro < 1.5$ for this system.}\\
\end{tabular}
\end{center}
\end{minipage}
\end{table*}

\begin{figure*}[t]
% \centering\input{figs/models.tex}
\centering\epsfig{file=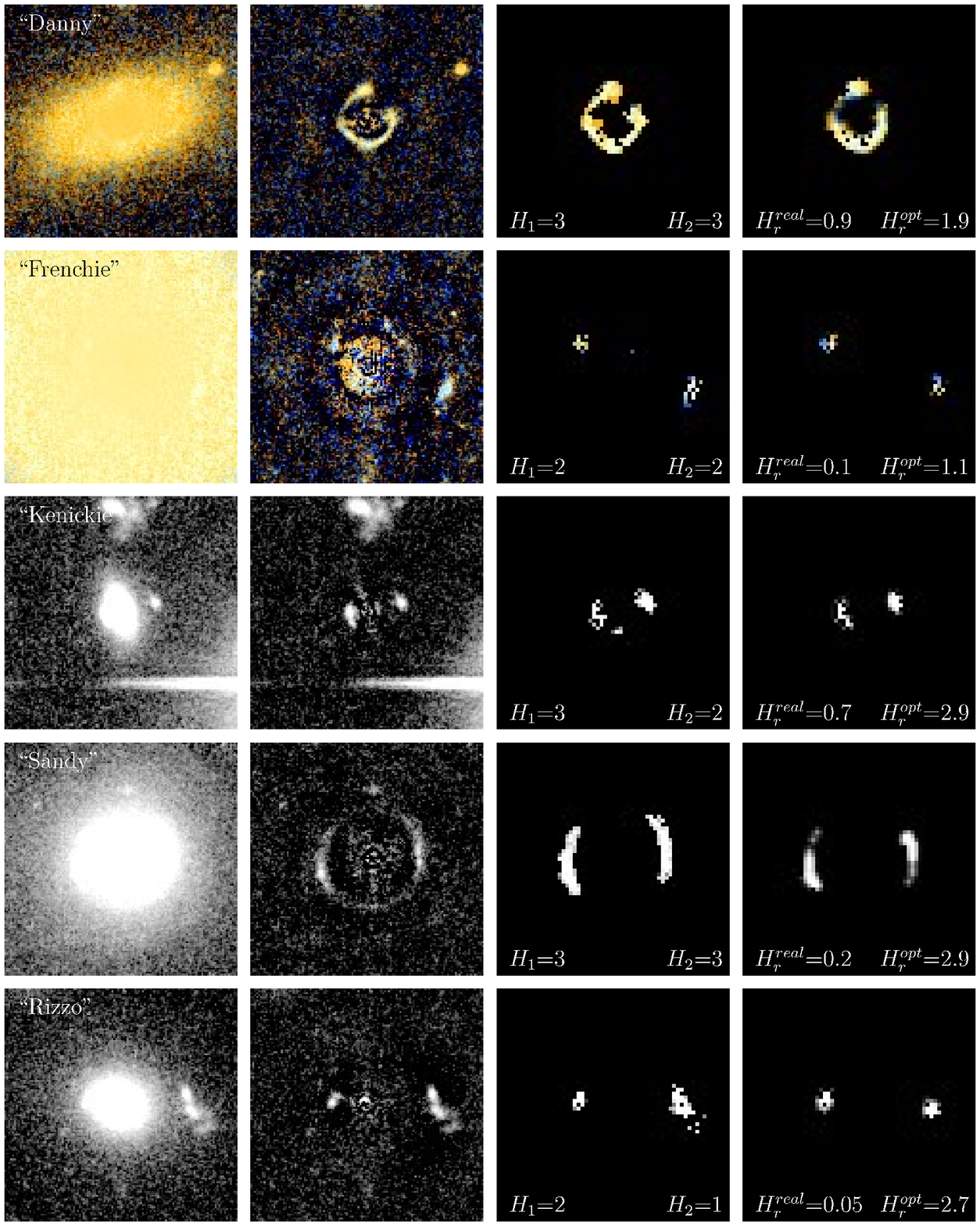,width=0.9\linewidth}
\caption{Lenses and lens models for new systems created by the HAGGLeS
robot.  We show the galaxy cutout (column 1), the subtraction residual
(column 2), the masked robot input image (column 3), and the
reconstructed images from the minimum source
required to produce the configuration  (column 4). 
The robot uses Moffat profile subtractions by
default; here we have improved upon this by using a B-spline
subtraction \citep[]{Bol++06}.  Cutouts are 6 arcseconds on a side.}
\label{fig:lensmodels}
\end{figure*}

% - - - - - - - - - - - - - - - - - - - - - - - - - - - - - - - - - - - - - - - - 

\subsection{Improved Robotic Lens Models}

As shown in Section~\ref{sec:accuracy}, and noted by \citet{Mar++08},
the lens model parameters returned by the HAGGLeS robot are not always
accurate. We find that the principal cause of robot modeling (and
indeed classification) error is insufficient lens light
subtraction. Disks and irregular profile slopes both give rise to
significant symmetrical residuals that confuse the robot. For the
small number of high quality lens candidates identified in
section~\ref{sec:results}, we can solve this problem on a case by case
basis, and provide the robot with cleaner images to model, and thus
produce more accurate estimates of the lens candidates' Einstein radii.

We improved the lens galaxy light subtraction with the flexible
B-spline fitting approach developed by \citet{Bol++06}, after first
masking out the objects in the image identified as candidate
lensed arcs. This procedure leaves sharper lensed image residuals. We
then set all undetected pixels to zero as described in
\citet{Mar++08}, but also at this stage masked out all the features in
the B-spline residual map not identified as lensed arcs. We then
re-ran the lens-modeling part of the HAGGLeS robot, and took as our
final estimated Einstein radius the position of the peak of the Gaussian fit
to the source plane flux curve, as described in \citet{Mar++08}.  The
resulting lens models and their parameters are shown in
Figure~\ref{fig:lensmodels} and Table~\ref{tab:lenspars}.

We can use our results from \S~\ref{subsec:proc1a} to estimate the
accuracy of the Einstein Radii measured by the robot. As shown in
Figure~ \ref{fig:cfrrein}, we found that -- for well modeled lenses
-- robot- and SLACS-measured Einstein radii agree to within
10\%. However, this can be considered an upper limit to our true
uncertainty, since we use improved subtractions for the new
systems. In practice, the robot's estimates using improved
subtractions are as good as the relatively simple models allow them to
be. To be conservative we adopt an error of 5\% on the robot's
Einstein Radii.

\begin{table*}
\begin{center}
\caption{Relevant lensing data for new lenses}
\label{tab:lenspars}
\begin{tabular}{cccccccccc}
\hline \hline
Short name&               Separation    & $\thetaE$     & F814W & R$_{\rm e}$ & $\zd$         & $\zs$                  & $\sigstar$    & $\sigma_{\rm *,FP}$ & Morphology  \\ 
          &               (arcsec)      & (arcsec)      &       & (arcsec)    &               &                        & (km s$^{-1}$) & (km s$^{-1}$)       &             \\\hline
Danny&    \multirow{2}{*}{107.9$\pm$0.2}& 0.65$\pm$0.03 & 18.75 & 0.61        & 0.34$\pm$0.03 & 0.48$^{+0.03}_{-0.02}$ & --            & 296$\pm$41          & S0          \\
Main lens &                             & 1.00          & 17.04 & 2.82        & 0.3223        & 0.5812                 & 358$\pm$17    & 313$\pm$43          & E/S0        \\
\hline							                   
Frenchie& \multirow{2}{*}{46.8$\pm$0.2} & 1.45$\pm$0.07 & 16.10 & 1.79        & 0.104         & 0.91$^{+0.15}_{-0.56}$ & 248$\pm$17    & 226$\pm$31          & E           \\
Main lens &                             & 1.02          & 14.96 & 4.80        & 0.106         & 0.4019                 & 279$\pm$13    & 269$\pm$37          & E/S0        \\
\hline							                   
Kenickie& \multirow{2}{*}{35.2$\pm$0.2} & 0.65$\pm$0.03 & 19.99 & 0.43        & 0.33$\pm$0.08 & 1.44$^{+0.22}_{-0.57}$ & --            & 162$\pm$22          & S0/Sa       \\
Main lens &                             & 1.42          & 15.60 & 3.23        & 0.123         & 0.6402                 & 308$\pm$15    & 253$\pm$35          & E/S0        \\
\hline							                   
Sandy&    \multirow{2}{*}{82.7$\pm$0.2} & 1.31$\pm$0.07 & 18.27 & 1.55        & 0.32$\pm$0.01 & 1.38$^{+0.21}_{-0.56}$ & --            & 233$\pm$32          & E           \\
Main lens &                             & 1.52          & 16.87 & 2.55        & 0.285         & 0.5753                 & 343$\pm$32    & 305$\pm$42          & Sa/S        \\
\hline							                   
Rizzo&    \multirow{2}{*}{106.0$\pm$0.2}& 1.24$\pm$0.06 & 19.47 & 0.46        & 0.41$\pm$0.03 & 1.12$^{+0.12}_{-0.37}$ & --            & 279$\pm$39          & E           \\
Main lens &                             & 1.68          & 16.43 & 1.95        & 0.158         & 0.7353                 & 211$\pm$12    & 247$\pm$34          & undetermined\\
\hline
\multicolumn{9}{l}{\scriptsize\raggedright Notes: F814W magnitudes are not corrected for Galactic
Extinction. Uncertainties are typically 0.03 on }\\
\multicolumn{9}{l}{\scriptsize\raggedright F814W total
magnitudes, 5\% on R$_{\rm e}$, and 2\% on SLACS Einstein Radii
\citep{Bol++08a}.}\\
\end{tabular}

%108.2$\pm$0.1
%47.2$\pm$0.1
%34.9$\pm$0.1
%82.9$\pm$0.1
%106.4$\pm$0.1
\end{center}
\end{table*}

% - - - - - - - - - - - - - - - - - - - - - - - - - - - - - - - - - - - - - - - - 

\subsection{Photometry}

Available from SDSS\footnote[1]{http://cas.sdss.org/dr6/en/} are
photometric redshifts and apparent lens galaxy magnitudes; with the
exception of one (Frenchie), no spectra are available. To supplement the
SDSS data, we use the {\tt galfit} software \citep{Pen++02} to fit de 
Vaucouleurs models
to the Hubble F814W data and derive apparent magnitude and
circularized effective radii, listed in
Table~\ref{tab:lenspars}. After correcting for Galactic extinction, we
calculate the rest frame V-band magnitude from the observed F814W, a
conversion for which there is very little scatter between different
spectral types.  Details on surface photometry and K-color corrections
can be found in the paper by \citet{Tre++01b}. For the main SLACS
lenses we use data from \citet[]{Bol++08a}.

% - - - - - - - - - - - - - - - - - - - - - - - - - - - - - - - - - - - - - - - - 

\subsection{Velocity dispersion and source redshift}

In this Section we use available spectroscopy and photometry to
estimate the velocity dispersion and source redshifts of the new lens
systems, using two physically motivated assumptions: i) early-type
lens galaxies lie on the fundamental plane \citep[hereafter FP][]{Dre++87,D+D87,Tre++06,Bol++08b}; ii) the ratio between stellar
velocity dispersion and that of the best fitting SIS is approximately constant.

We use the photometric redshift and evolution-corrected V band
luminosities to estimate the central stellar velocity dispersion
predicted by the FP relation:

\begin{equation}
\log R_{\rm e}  = a \log \sigma_{\rm e,2} + b \log I_e + c,
\label{eq:FP}
\end{equation}

\noindent
where $\sigma_{\rm e,2}$ is the stellar velocity dispersion corrected
to an aperture of radius half the effective radius (in units of 100
kms$^{-1}$), I$_{\rm e}$ is the effective surface brightness in units
of 10$^9$ L$_{\odot}$ kpc$^{-2}$, and R$_{\rm e}$ is the effective
radius in kpc. We adopt the coefficients $a=1.28$ $b=-0.77$ $c=-0.09$
derived in SLACS-V for the SLACS sample. The intrinsic
scatter of the fundamental plane dominate the uncertainty on the
estimated $\sigma_{\rm e,2}$ (0.05 dex).  Central stellar velocity
dispersions $\sigstar$ are obtained from $\sigma_{\rm e,2}$ using the
standard correction described in SLACS-V.
      
Additionally, SLACS-IV and -V found that $\sigstar$ is
correlated with $\sigSIE$, the velocity dispersion that best fits
the model of the lens as a singular isothermal ellipse (SIE).  For the
SLACS sample:

\begin{equation}
\langle \sigstar / \sigSIE \rangle =1.02\pm0.01.
\label{eq:ff}
\end{equation}

We may also calculate $\sigSIE$, assuming the Einstein
radius $\thetaE$, lens redshift, and source redshift are known:

\begin{equation}
% \sigma_{\rm SIE} = \left(\frac{c}{{\rm km\ s}^{-1}}\right) \sqrt{\left(\frac{1}{4\pi}\frac{\theta_E}{{\rm radians}}\right)\frac{D_S}{D_{DS}}},
\sigSIE = c \sqrt{\frac{\thetaE}{4\pi}\frac{\Ds}{\Dds}},
\label{eq:sSIE}
\end{equation}

\noindent where $\Ds$ and~$\Dds$ are, respectively, angular-diameter 
distances to the source galaxy and between the lens and source
galaxies, and $\thetaE$ is given in radians.

We combine Equations~\ref{eq:FP} to~\ref{eq:sSIE} to determine our
best estimate of the source redshift. To obtain the posterior
probability distribution function of~$\zs$, we assume that $\sigstar$
is log-normally distributed with scatter 0.06~dex, which is dominated
by the intrinsic scatter of the FP and of Eq~\ref{eq:ff}. We adopt
priors appropriate for the source population.  For the newly
discovered -- imaging selected -- lenses, we adopt as prior the
redshift distribution of faint galaxies in single orbit ACS-F814W data
as measured by the COSMOS survey. For the main SLACS lenses we adopt
the same prior, but truncated at $\zs < 1.5$, i.e.\ the highest redshift where
[\ion{O}{2}] is still visible within the observed wavelength range
covered by the SDSS spectrograph used for discovery. The results
change very little if a uniform prior is adopted instead.

\begin{figure}
\centering\epsfig{file=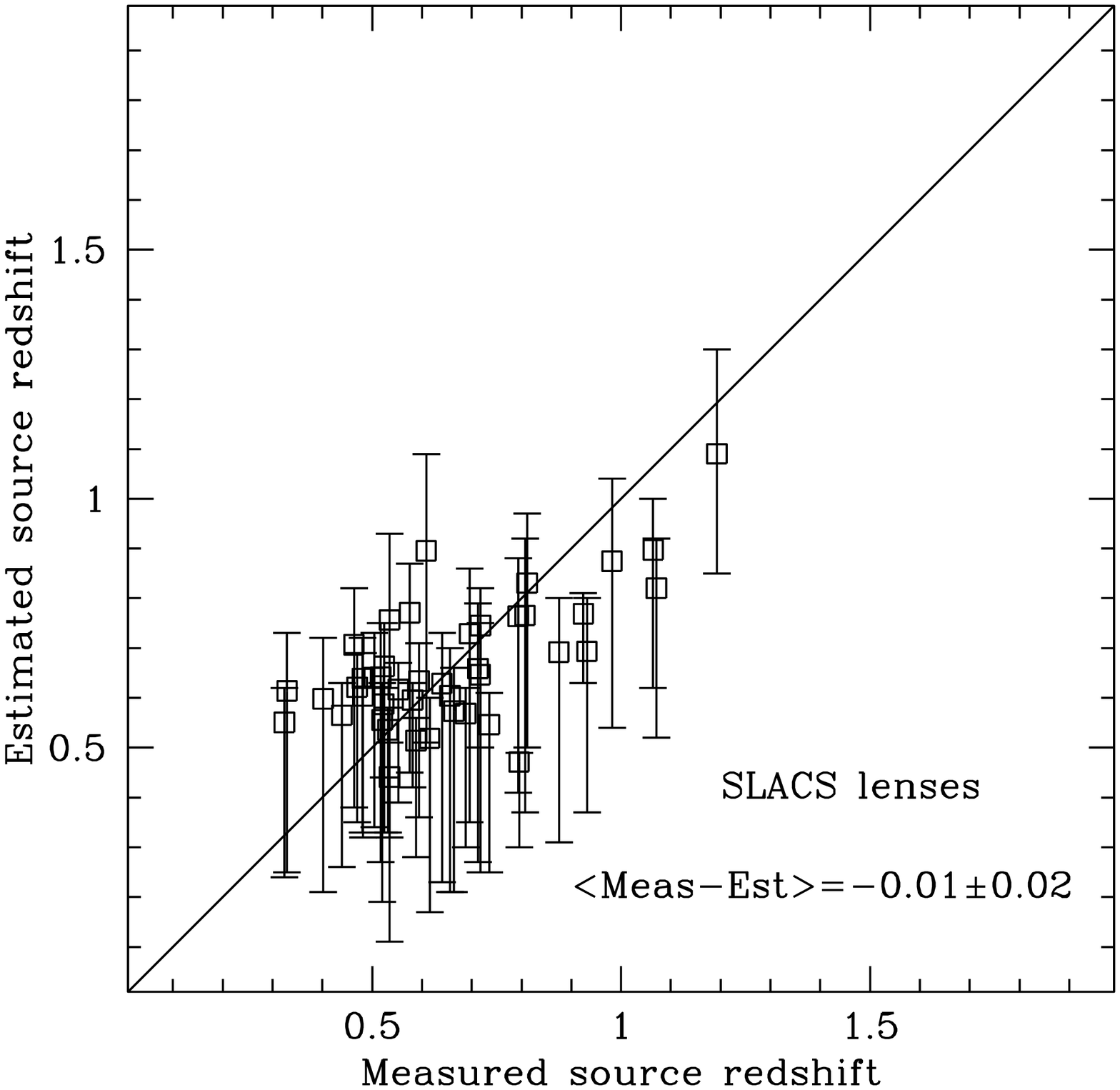,width=0.9\linewidth}
\caption{Source redshift estimated from the FP method vs. the corresponding spectroscopic source
redshifts for the SLACS lenses. }
\label{fig:cfrvzs}
\end{figure}

\begin{figure}
\centering\epsfig{file=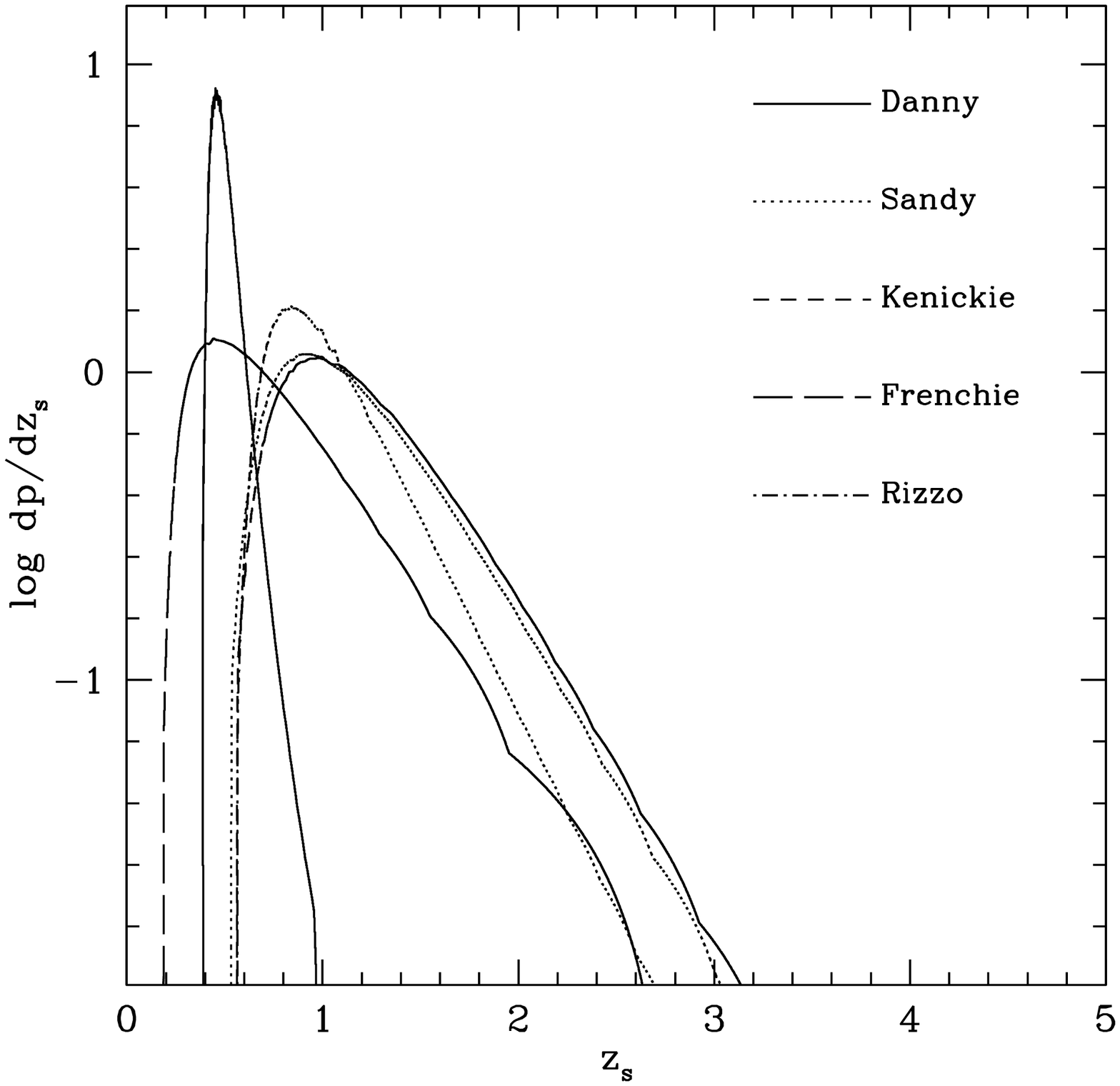,width=0.9\linewidth}
\caption{Source redshift posterior probability distribution function estimated for the newly discovered lenses using the FP method. }
\label{fig:zsnew}
\end{figure}

The result of this calculation for the main lenses can be seen in
Figures~\ref{fig:cfrvzs} and~\ref{fig:zsnew}. The former compares the
estimated source redshifts to the known source redshifts for the main
SLACS sample. This sanity check indicates that our procedure is unbiased
and that the scatter is consistent with the estimated error bars. The
latter figure shows the posterior probability distribution function
for the source redshifts of the newly discovered lenses. As expected,
the posterior is asymmetric with a tail to high-z due to to the strong
dependency on the ratio of angular diameter distances on the source
redshift, when it approaches the lens redshift.

The estimated stellar velocity dispersions and source redshifts are
given in Table \ref{tab:lenspars}.  We include the spectroscopically
measured $\sigma_*$ when available, for comparison.

% ----------------------------------------------------------------------------
 
\section{Discussion}\label{sec:discussion}

We begin this section by summarizing the arguments that led to our
classification of the five newly discovered lens systems in
\S~\ref{sec:validity}.  We then discuss the environments in which the
lenses are found in \S~\ref{sec:env}, and in \S~\ref{sec:compare} we
compare the properties of the new systems to the SLACS main lenses.

%  - - - - - - - - - - - - - - - - - - - - - - - - - - - - - - - - - - - - - -
 
\subsection{Validity and Properties of Candidates}\label{sec:validity}

Our final classification is entirely imaging-based (except in Frenchie's
case, where an SDSS spectrum was available).  
We thus rely entirely on lens geometry, the appearance of the 
subtraction residuals,
and the robot's ability to model the lens light. For this reason our
standards are high: we require clearly-identified multiple images in
all cases, and a straightforwardly-modeled 
image configuration. When available,
colors are used to strengthen the case. As an additional sanity check,
we note that the Einstein Radii are consistent with those expected
from the FP and a simple SIE model, for sensible values of source
redshifts (see Table~\ref{tab:lenspars}). We now discuss each case
individually.

Danny's lensing galaxy is a large red elliptical, while the lensed
source is blue, as expected for lens systems.   The identified quad geometry is a typical lens geometry
and may be called ``cusp dominated'' \citep[e.g.][]{Koc++06}, as it occurs when the source lies
near to a cusp of the inner caustic.  Although we also noted a strong
disk component remaining after the initial Moffat profile subtraction, the
robot is able to effectively model Danny as an SIS+external shear (a
situation in which such a geometry would occur).

Sandy has two strong arcs on either side of the lens galaxy, consistent
with a double pattern produced by a source almost directly behind a
SIS; this is well modeled by the robot.  
We note a fainter peak in surface brightness above the
lens galaxy; with no obvious counter image,
we believe this object is likely a small satellite galaxy in the lens plane.

The two images comprising Kenickie's lensing event, and the center
of the lensing galaxy, are not all perfectly aligned: this suggests that
either the source does not lie quite on the optical axis of an 
ellipsoidal lens or
that external shear is present.  We note that the inner
image appears to have more curvature than the outer; this could be 
brought about
by unsubtracted lens-plane structure that a color image would rule out.

Frenchie, another double, has been imaged in two filters and has colors 
consistent with lensing.  This is
the only one of our systems backed by spectroscopic data.  We find a
spectroscopic redshift of $\zd = 0.104$. Also, a stellar velocity
dispersion is available from SDSS, corresponding to
$\sigstar = 248\pm17$ km\ s$^{-1}$ after aperture correction, in good
agreement with the value estimated via the FP technique ($226\pm27$
km\ s$^{-1}$)

Rizzo is able to be modeled as a double; however the morphologies of the two
identifiable images are not as well-matched as in the previous 4 cases. In the
lensing scenario, the source would lie only partly within the outer caustic of
the lens, with the more extended outer image being only partially
strongly-lensed.

%  - - - - - - - - - - - - - - - - - - - - - - - - - - - - - - - - - - - - - -

\subsection{Lens environments}\label{sec:env}

We expect lenses, as massive galaxies, to be clustered. Consistent
with this hypothesis, 3/4 new lenses are found at redshifts very
similar to those of the main SLACS lens in the field. In this section
we study the environment of the fields with more than one lens to
identify possible large scale structures, using the environment
measures of local and global density as defined in SLACS-VIII.

SLACS-VIII found no significant difference between the environment of
SLACS lenses and that of non-lensing, but otherwise identical, galaxies.
With the exception of Frenchie, the environments of the newly
discovered lenses are fairly typical of the overall distribution of
SLACS lens environments, with the field of Kenickie being slightly
under-dense, while that of Sandy being somewhat over-dense. Frenchie, as
previously mentioned, lies in a very over-dense environment, the
densest of all SLACS fields.

%They define $\Sigma_{10}$ as the surface density
%of galaxies within the tenth nearest neighbor, and $D_1$ as the
%density of galaxies within a radius $h^{-1}$.  The ``local'' and
%``global'' over-densities are the normalized $\Sigma_{10}$ and $D_1$,
%respectively. Values determined for the main lenses of each our five
%systems are shown in Table \ref{tab:env} \citep[after][]{Tre++08}. The
%average local density for all SLACS lenses was determined to be
%2.90$\pm$1.20 while the average global density was 1.10$\pm$0.09.

In fact, both Frenchie and Sandy appear associated with known
clusters.  Frenchie lies within SDSS-C41035, at redshift 0.106, which
places both Frenchie ($\zd = 0.104$) and its main lens ($\zd = 0.106$) as
members.  The cluster in Sandy's field, MaxBCGJ217.49493+41.10435, is
at redshift 0.270, and the main lens at $\zd = 0.285$ lies within the
cluster.  We have only photometric redshift for Sandy ($\zd = 0.32\pm0.01$), 
which places it slightly beyond the extent of the
cluster; however given the possible systematic errors in 
the photometric redshift we
deem it likely that Sandy is also a member of the field's cluster.
 
\begin{table*}
\caption{Environments of new lenses and candidates}
\label{tab:env}
\begin{tabular}{cccccc}
\hline \hline
Short name&SDSS field&local overdensity&global overdensity&Cluster&$z_{Cl}$\\  \hline
Danny& J0737+3216& 1.55$\pm$0.50&1.22$\pm$0.28&None\\
Frenchie& J1143-0144& 79$\pm$25.5&3.7$\pm$0.50&SDSS-C41035&0.106\\
Kenickie& J1213+6708& 0.48$\pm$0.16& 0.45$\pm$0.16&None \\
Sandy& J1430+4105& 1.83$\pm$0.62& 1.22$\pm$0.34&MaxBCGJ217.49493+41.10435&0.270\\
Rizzo& J1103+5322& 0.98$\pm$0.3&0.75$\pm$0.18&None\\
\hline
\end{tabular}
\end{table*}

In conclusion, the results suggest that lens fields associate with
galaxy over-densities are the most likely to present additional strong
lens phenomena. Only $12/70 = 17\%$ of all SLACS lenses are associated with
known clusters, but $50\%$ of the newly discovered ones are (2/5 when
including
the likely candidate).  With the present data it is hard to
disentangle the contribution to the boost in strong lens surface
density by the enhanced surface density of deflectors.
%from the extra convergence made available by cluster and group scale halos.

%  - - - - - - - - - - - - - - - - - - - - - - - - - - - - - - - - - - - - - -
 
\subsection{Comparison to SLACS lenses}\label{sec:compare}

\begin{figure}
\centering\epsfig{file=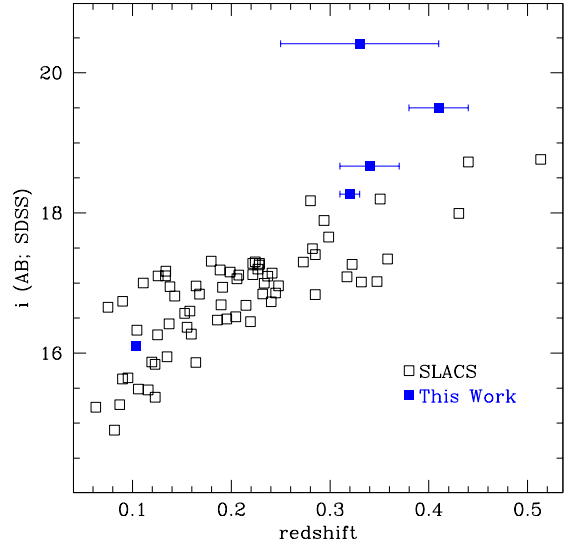, width=0.9\linewidth}
\caption{Plot of apparent i-band magnitude vs.\ lens redshift for the
SLACS main lenses and new systems.  We find that four of the five new
systems are less luminous than the main lenses.}
\label{fig:plotiz}
\end{figure}

\begin{figure}
\centering\epsfig{file=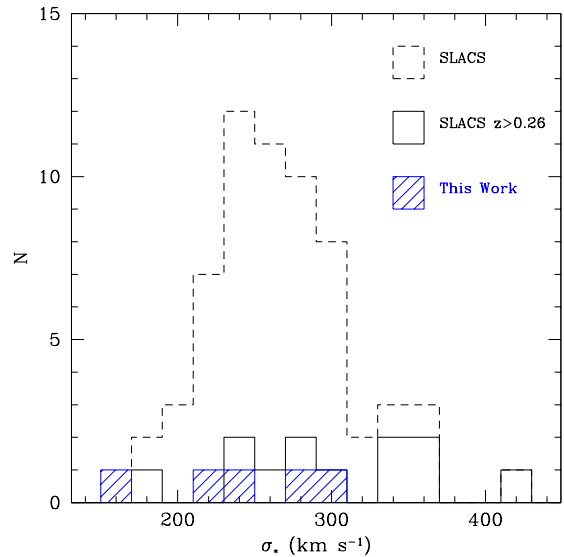, width=0.9\linewidth}
\caption{Histogram of central velocity dispersions for main lenses and new systems; new systems are shaded.  Main lens velocity dispersion were aperture-corrected from SDSS photometry, while we used the FP relation to calculate the values for new lenses.}
\label{fig:historsig}
\end{figure}

We anticipated discovering lensing events similar to, but fainter
than, the SLACS main lenses; we find this to be largely true.  In
Figure~\ref{fig:plotiz}, we plot apparent $i'$ magnitude against
redshift for the main lenses and new systems.  We find that four of
the new systems are less luminous than the SLACS lenses at the same redshift,
but that Frenchie is comparable.  We show the distribution of stellar
velocity dispersions in Figure~\ref{fig:historsig}; new systems are
shaded.  

To interpret these histograms we need to take into account
the redshift dependence of the properties of the SLACS main sample. At
the lower redshifts, the sample is dominated by the more abundant,
slightly less massive galaxies. At higher redshifts, the flux limit of
the SDSS spectroscopic database leaves only the most massive
galaxies. Keeping this in mind, we also plot the distributions for the
SLACS sample in the same redshift range of all the new lenses save Frenchie ($>0.26$ is chosen because of a natural break in the
redshift distribution; the choice of threshold does not influence our
conclusions). Indeed, the newly identified lenses are less
massive than the SLACS main lenses when both samples are restricted to
$\zd > 0.26$. The average $\sigma_*$ for the two samples are
respectively $243\pm35$ km\ s$^{-1}$ and $299\pm20$ km\ s$^{-1}$.

% - - - - - - - - - - - - - - - - - - - - - - - - - - - - - - - - - - - - - - - - 

\subsection{Lens abundance}

In Table \ref{tab:rates}, we compare lensing rates (or abundances, in
degree$^{-2}$) for four HST imaging surveys, including this one. The
quoted uncertainties on the inferred rates delimit the Bayesian 68\%
confidence interval, assuming Poisson statistics and a uniform prior
PDF.  As noted previously, classification systems across surveys are
not consistent, thus one must impose similar criteria when comparing
the results of different surveys; naturally there will be differences
in opinion.  In this work we do not rely on spectroscopic data to
classify a lens as ``definite'' and instead require that lens
morphology, surface brightness, and color (if available) are clearly
identifiable with a typical lens geometry.  Due to the limited data
available, we apply these criteria rigorously.  With the goal of
applying similar criteria to all surveys considered, we take as
definite lenses: the literature-confirmed lenses for MDS, the
``unambiguous'' candidates for AEGIS, the ``best'' candidates for
COSMOS, and the $H = 3$ lenses for this work.

The largest survey of ``blank'' sky undertaken to date is the
COSMOS survey \citep{Fau++08}, whose findings imply a measured lensing
rate of around $12$ lenses per square degree. Since our data are of
comparable depth (and in the same single filter), we adopt this as our
fiducial value.  Given this lensing rate, we might expect to find 1.61
lenses, instead of our observed 4. Assuming a uniform prior on the
lensing rate we find that the inferred lensing rate from our survey is
$30^{+24}_{-8}$ lenses/degree$^2$.

The uniform prior is somewhat unrealistic -- it does not down-weight
the occasional high lens yields that could arise as statistical flukes
from the long-tailed Poisson likelihood. A maximally conservative approach 
would be to take the
COSMOS rate as the mean of an exponential prior; in this case, we infer a lensing rate of
$18^{+14}_{-5}$ lenses/degree$^2$.  We still find a
significantly higher lens abundance than seen in the COSMOS survey:
with the COSMOS prior the probability that the lensing rate in the
SLACS fields is greater than 12~degree$^{-2}$ is 88\%. Relaxing the 
assumption that our fields are similar in nature to those in
COSMOS and reverting to the uniform prior, 
we find that there is only a $\sim2\%$ chance that the lensing rate
in the SLACS fields is less than the COSMOS rate of 12 lenses/degree$^2$. 

%The uniform prior is somewhat unrealistic -- it does not down-weight
%the occasional high lens yields that could arise as statistical flukes
%from the long-tailed Poisson likelihood. It also ignores the
%experience in the previous lens surveys, and all other astrophysical
%information.  A maximally conservative approach would be to take the
%COSMOS rate as the mean of an exponential prior (appropriate for
%situations where the mean of the prior distribution is known, but
%nothing else). In this case, we infer a lensing rate of
%$18^{+14}_{-5}$ lenses/degree$^2$. Even refusing to acknowledge the
%fact that elliptical galaxies are clustered, and assuming that our
%fields are comparable to those in the COSMOS survey, we still infer a
%significantly higher lens abundance than seen in the COSMOS survey:
%with the COSMOS prior the probability that the lensing rate in the
%SLACS fields is greater than 12~degree$^{-2}$ is 88\%.
% The lensing rate for this 
% work, at 30 lenses/degree$^2$, is the highest of the high resolution imaging  
% surveys considered. 
% We additionally find that,  

Although no definitive conclusions can be drawn due to the small numbers and
variety of surveys, the results from this project do support the hypothesis of
\citet{Fas++06}
that looking near known lenses increases the efficiency of finding new
lenses.  
% 
% Support for this idea also comes from the serendipitous discovery of
% two new lenses within a single ACS pointing, in the field of the known lens
% B1608 \citep[an area of $\sim$11 sq. arcmin][]{Fas++06}.

% We calculate a lensing rate of 720
% lenses/degree$^2$ for this field, with an upper limit of 2130 (if both new
% systems are lenses), and a lower limit of 490 (if only one new system is a
% lens). 
%Although the lensing rate in this one field certainly cannot be
%directly compared to those from dedicated lens searches, we consider it
%additional evidence for the clustering of lenses.

%\begin{figure}
%\centering\epsfig{file=figs/historein.eps, width=\linewidth}
%\caption{Histogram of Einstein radii for main lenses and new lenses; new systems are shaded.  Einstein radii for main lenses are from the SLACS papers; those for the new systems were calculated by the robot.}
%\label{fig:historein}
%\end{figure}

\begin{table}
\begin{center}
\caption{Lensing rates for selected surveys}
\label{tab:rates}
\begin{tabular}{lccc}
\hline\hline
Survey            & Area       & $N_{\rm lenses}$ & Lensing rate    \\ 
                  & (deg$^2$)  &                  & (degree$^{-2}$) \\ \hline
MDS$^1$           & $0.17$     &        2         & $12^{+15}_{-3}$ \\     
AEGIS$^2$         & $0.19$     &        3         & $16^{+15}_{-5}$ \\     
COSMOS$^3$        & $1.64$     &       20         & $12^{+4}_{-2}$  \\     
\hline
This Work         & $0.13$     &        4         & $31^{+24}_{-9}$ \\    
with COSMOS prior &            & (predicts 1.6)   & $18^{+14}_{-5}$ \\  \hline  
\end{tabular}\end{center}
Confidence intervals are Bayesian 68\%, assuming Poisson statistics and a
uniform prior on the lensing rate, except in the final row where the COSMOS
rate is taken as the mean of an exponential prior. Relevant citations are as
follows:
$^1$ \citet{Rat++99},
$^2$ \citet{Mou++07},
$^3$ \citet{Fau++08}
\end{table}

% ----------------------------------------------------------------------------

\section{Conclusions}\label{sec:conclusions}

We have performed a highly efficient search for gravitational lenses
based purely on imaging data.  Our search area consisted of 44 HST/
ACS fields, each centered on a SLACS ``definite'' lens, therefore
exploiting the expected clustering of gravitational lenses in each
field.  We compared the purity, completeness, and investigator time
for four different search methods.  These methods are comprised of:
(1) use of the output from the HAGGLeS robot, (2) inspection of BG
subtraction residual images, (3) inspection of BG cutout images, (4)
full frame visual inspection of ACS fields.

Our main conclusions are:

\begin{itemize}

\item Taking into account efficiency as well as completeness and purity, we
find that of the methods used, procedures 1 (using output from the
HAGGLeS robot) and 2 (looking at subtraction residuals of bright galaxies) have the best performance.  
In situations where the simple SIS+external shear model
used by the robot may be insufficient -- such as in clusters or to
find naked cusp configurations -- it may be most effective to use a
combination of procedures 1 and 2, in order to inspect all objects
modeled by the robot.  However, looking at only the bright galaxies (BGs) in
the fields
did not decrease the completeness of this particular survey, while
doing so greatly improved efficiency.

\item We discovered four new strong lenses and one promising candidate in the
course of our survey.  We find that 3/4 of these new lens systems have
lens redshifts similar to those of their main lenses; additionally,
two of the new lenses are found in clusters of galaxies 
that also include their respective main lenses.

\item We find that 3/4 new systems are less luminous and 
less massive
than the SLACS lenses.  Overall, we are probing early-type galaxies at
higher redshifts and lower masses than the SLACS survey. For these
comparisons, we used the data from the improved robot models, available
photometry, and the Fundamental Plane to estimate the central velocity
dispersions and source redshifts for each of our new systems.

\item The lens abundance for this survey, $30^{+24}_{-8}$ lenses/degree$^2$ 
(uniform prior), is markedly higher than the lensing rates for the three other
HST surveys considered at comparable depths and resolution.  Despite
the small numbers and variations in search methods, this result
supports the idea that searching near known lenses increases the yield
of a lens survey.

\end{itemize}

The HAGGLeS project is currently using a combination of procedures 1
and 2 to search for new lenses in the HST archive; through efforts
such as this and others, we will be able to refine our lens search
techniques for future surveys covering much larger areas of the sky.
The use of efficient and repeatable lens search methods will further
us towards the goal of having a large, homogenous sample of strong gravitational
lenses. Such a sample will enable us to calculate a lens-lens correlation
function and constrain the statistical properties of halos containing
lens galaxies.

% ----------------------------------------------------------------------------

\acknowledgments

This paper builds on the work of the SLACS and HAGGLES
collaborations. We are grateful to our SLACS and HAGGLES collaborators
and friends -- Adam Bolton, Roger Blandford, Maru\v{s}a Brada\v{c},
Scott Burles, Chris Fassnacht, Raph\"{ae}l Gavazzi, David Hogg, Leon
Koopmans, Leonidas Moustakas, Eric Morganson, and Tim Schrabback-Krahe
-- for their many insightful comments and suggestions throughout this
project.  Support for HST programs
\#10174, \#10587, \#10886, \#10676, \#10494, and \#10798 was
provided by NASA through a grant from the Space Telescope Science
Institute, which is operated by the Association of Universities for
Research in Astronomy, Inc., under NASA contract NAS 5-26555.  
E.R.N.\ acknowledges partial financial support from the College of Creative Studies.
The work of P.J.M.\  was
supported by the TABASGO foundation in the form of
a research fellowship.
T.T.\ acknowledges support from the NSF through CAREER award NSF-0642621, by
the Sloan Foundation through a Sloan Research Fellowship, and by the
Packard Foundation through a Packard Fellowship.  This research has
made use of the NASA/IPAC Extragalactic Database (NED) which is
operated by the Jet Propulsion Laboratory, California Institute of
Technology, under contract with the National Aeronautics and Space
Administration. This project would not have been feasible without the
extensive and accurate database provided by the Digital Sloan Sky
Survey (SDSS).  Funding for the creation and distribution of the SDSS
Archive has been provided by the Alfred P. Sloan Foundation, the
Participating Institutions, the National Aeronautics and Space
Administration, the National Science Foundation, the U.S. Department
of Energy, the Japanese Monbukagakusho, and the Max Planck
Society. The SDSS Web site is http://www.sdss.org/.  The SDSS is
managed by the Astrophysical Research Consortium (ARC) for the
Participating Institutions. The Participating Institutions are The
University of Chicago, Fermilab, the Institute for Advanced Study, the
Japan Participation Group, The Johns Hopkins University, the Korean
Scientist Group, Los Alamos National Laboratory, the
Max-Planck-Institute for Astronomy (MPIA), the Max-Planck-Institute
for Astrophysics (MPA), New Mexico State University, University of
Pittsburgh, University of Portsmouth, Princeton University, the United
States Naval Observatory, and the University of Washington.

% ----------------------------------------------------------------------------

\clearpage

\appendix

\section{Supplementary Figures and Tables}

\begin{figure*}[t]
\centering\epsfig{file=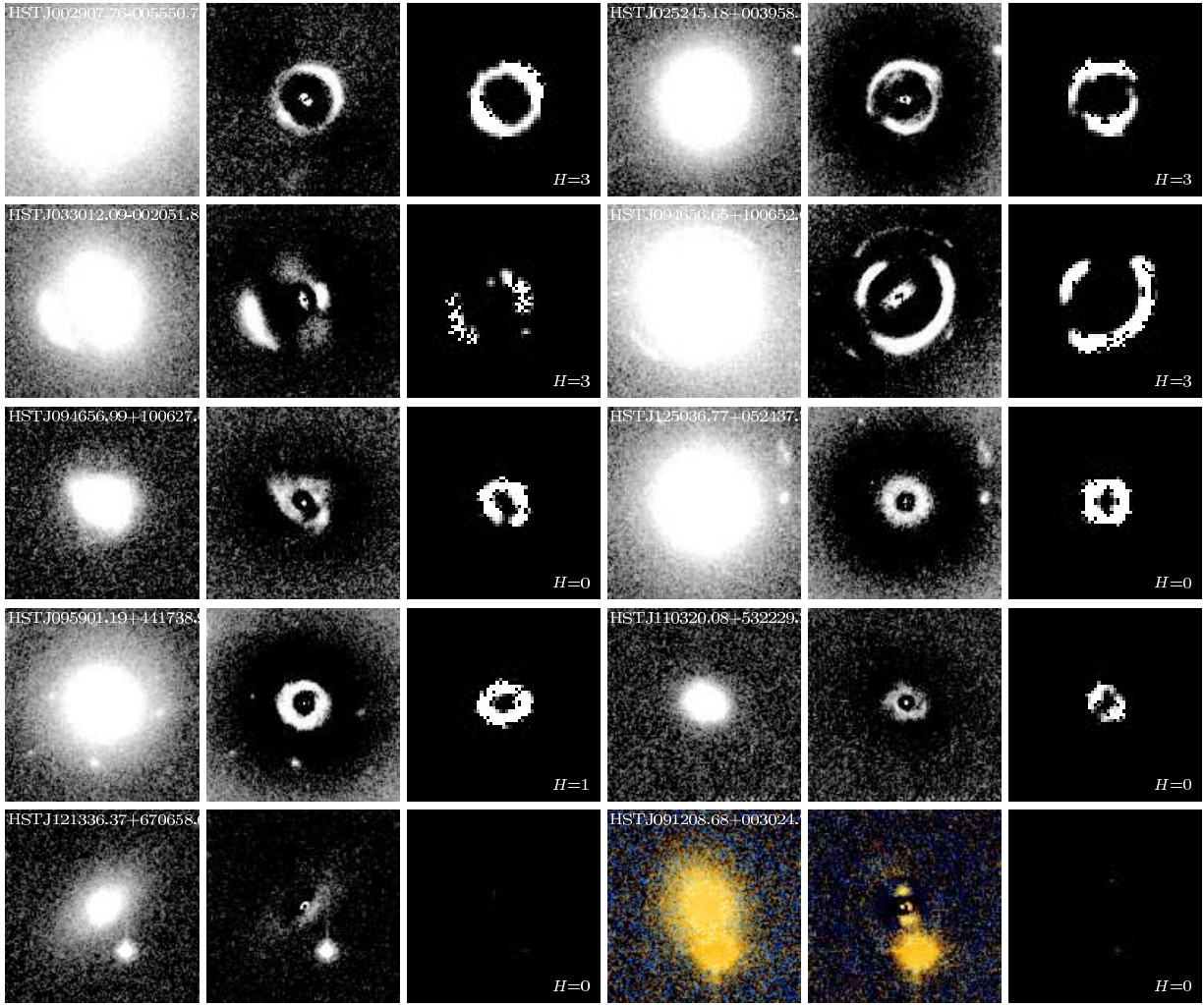,width=0.9\linewidth}
\caption{Objects classified by the robot as $\hrr > 2.5$, including four main 
lenses (first and second rows).  Cutouts are 6 arcseconds on a side.  
We find common false identifications to be rings left by the Moffat 
subtraction, unmasked disks, and nearby galaxies.\label{fig:classA}}
\end{figure*}

In this Appendix we give a more complete illustration of the HAGGLeS robot's
performance when given the SLACS main lenses. In Figure~\ref{fig:classA} we
show all objects classified by the \emph{realistic} robot as $\hrr > 2.5$:
\citet{Mar++08} found this to give a sample with $\sim20\%$~completeness  but
$\sim100\%$ purity. In the SLACS fields, 10 objects were classified as
$\hrr > 2.5$, including 4 main lenses. Comparing to the human-classified results
using the same input data (Procedure 2, Table~\ref{tab:stats}), this
represents a completeness of $~\sim11\%$ and a purity of $40\%$. Some
explanation for these differences are given in the main text. 

\begin{figure*}[t]
\centering\epsfig{file=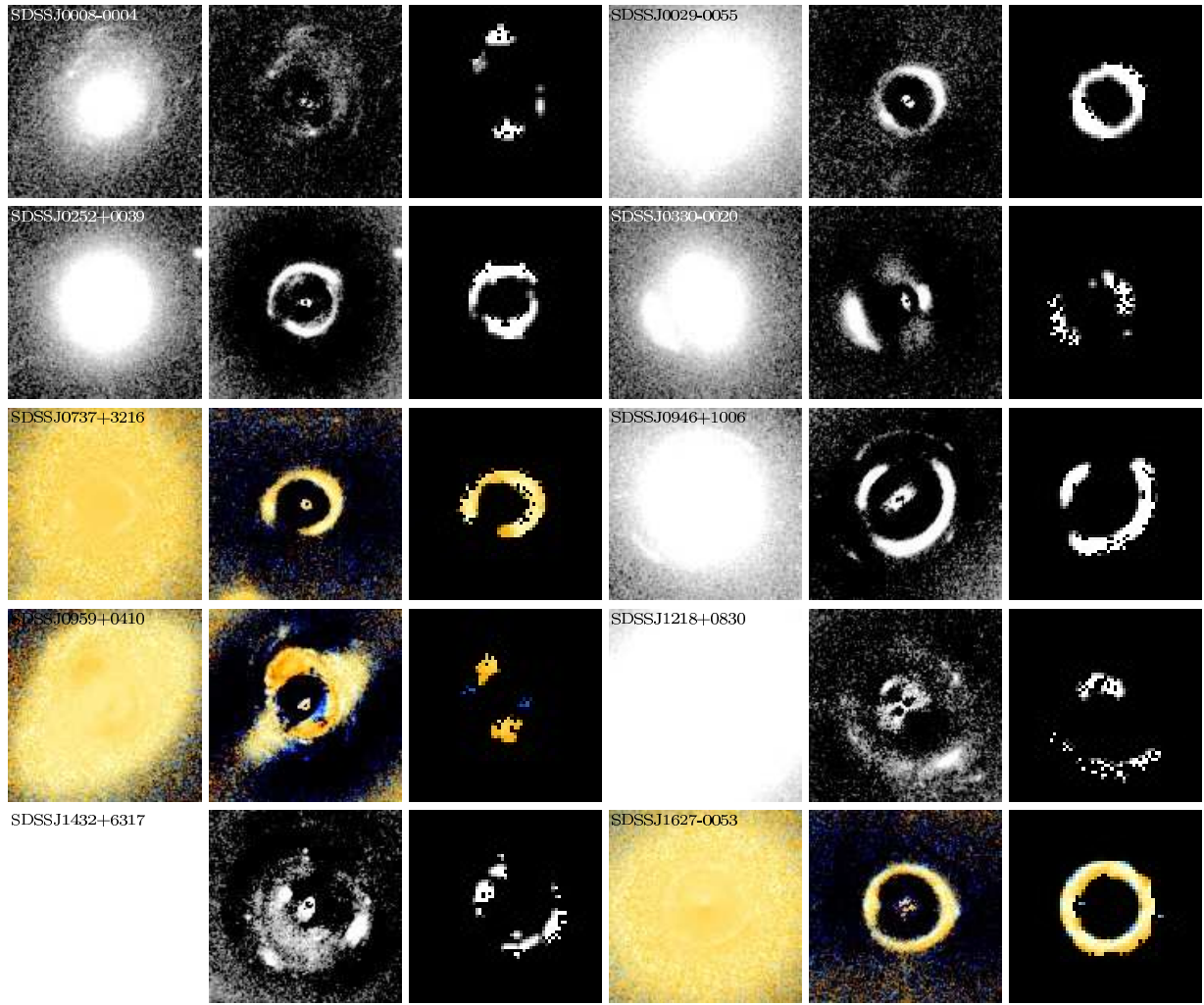,width=0.9\linewidth}
\caption{Robot output for SLACS main lenses with $\hro > 2.5$, including: 
galaxy cutouts (col. 1,4), residuals after Moffat subtraction (col. 2,5), 
and predicted images from the robot lens models (col. 3,6). 
In this figure and the two following, cutouts are 6 arcsec on a side.\label{fig:mainlensesA}}
\end{figure*}

\begin{figure*}[t]
\centering\epsfig{file=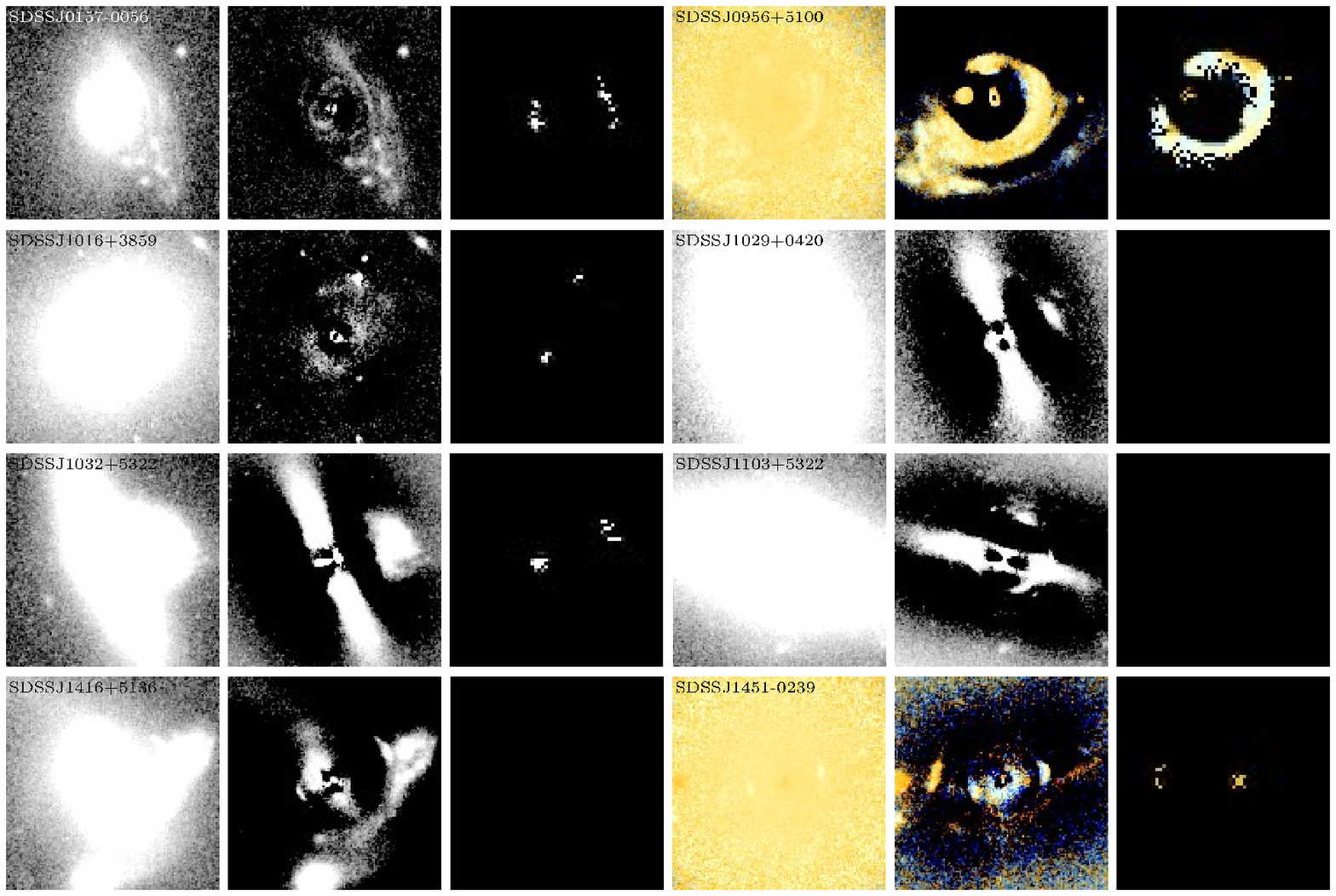,width=0.9\linewidth}
\caption{Robot output for SLACS main lenses with $\hro < 1.5$, including 
galaxy cutouts, subtraction residuals, and predicted images 
from the robot lens models.\label{fig:mainlensesX}}  
\end{figure*}

The classification of the SLACS main lenses is listed in full in
Table~\ref{tab:mainlenses}. The survey cutout images of these systems, sorted
into bins in $\hro$, are shown in Figure~\ref{fig:mainlensesA} ($\hro > 2.5$), 
Figure~\ref{fig:mainlensesB} ($1.5 < \hro < 2.5$), and 
Figure~\ref{fig:mainlensesX} ($\hro < 1.5$). For each system, we show the full
cutout image as presented for inspection in Procedure~3, the lens
galaxy-subtracted cutout image as presented for inspection in Procedure~2, and
the lensed images and counter-images predicted by the HAGGLeS robot's best
lens model.

\begin{table*}
\begin{center}
\caption{Classification of SLACS main lenses by method}
\label{tab:mainlenses}
\begin{tabular}{cccccccc}
\hline\hline
Object name    & \hon & \htw & \hth & \hfo & $\hro$ & $\hrr$ & Additional  \\
               &      &      &      &      &      &      & ACS filters \\ \hline
SDSSJ0008-0004 &   2  &   2  &   2  &   3  &  2.9 &  0.7 &    -- \\
SDSSJ0029-0055 &   3  &   3  &   0  &   3  &  2.9 &  2.8 &    -- \\
SDSSJ0157-0056 &   0  &   3  &   1  &   3  &  1.3 &  0.0 &    -- \\
SDSSJ0216-0813 &   3  &   3  &   0  &   3  &  2.0 &  1.9 &    -- \\
SDSSJ0252+0039 &   3  &   3  &   0  &   3  &  2.9 &  2.7 &    -- \\
SDSSJ0330-0020 &   3  &   3  &   2  &   3  &  2.9 &  2.8 &    -- \\
SDSSJ0728+3835 &   3  &   3  &   0  &   3  &  1.9 &  1.9 &    -- \\
SDSSJ0737+3216 &   3  &   3  &   3  &   3  &  2.9 &  2.4 & F555W \\
SDSSJ0808+4706 &   3  &   3  &   0  &   3  &  1.9 &  0.9 &    -- \\
SDSSJ0822+2652 &   3  &   3  &   0  &   3  &  2.4 &  1.9 &    -- \\
SDSSJ0841+3824 &   2  &   2  &   0  &   3  &  2.0 &  1.3 &    -- \\
SDSSJ0903+4116 &   3  &   3  &   1  &   3  &  1.9 &  1.7 &    -- \\
SDSSJ0912+0029 &   3  &   3  &   0  &   2  &  1.8 &  0.1 & F555W \\
SDSSJ0936+0913 &   3  &   3  &   0  &   2  &  1.9 &  1.3 &    -- \\
SDSSJ0946+1006 &   3  &   3  &   2  &   3  &  2.9 &  2.8 &    -- \\
SDSSJ0956+5100 &   0  &   3  &   1  &   3  &  0   &  0   & F555W \\
SDSSJ0959+0410 &   3  &   3  &   1  &   3  &  2.0 &  1.9 & F555W \\
SDSSJ0959+4416 &   2  &   1  &   0  &   3  &  2.9 &  0.7 &    -- \\
SDSSJ1016+3859 &   0  &   1  &   1  &   2  &  1.0 &  0.0 &    -- \\
SDSSJ1020+1122 &   3  &   3  &   1  &   2  &  1.5 &  0.8 &    -- \\
SDSSJ1023+4230 &   3  &   3  &   0  &   3  &  1.9 &  1.7 &    -- \\
SDSSJ1029+0420 &   0  &   1  &   0  &   1  &  0   &  0   &    -- \\
SDSSJ1032+5322 &   0  &   1  &   0  &   1  &  1.4 &  0.2 &    -- \\
SDSSJ1103+5322 &   0  &   3  &   0  &   3  &  0   &  0   &    -- \\
SDSSJ1142+1001 &   2  &   2  &   0  &   2  &  1.9 &  1.9 &    -- \\
SDSSJ1143-0144 &   3  &   3  &   2  &   3  &  1.9 &  0.5 & F555W \\
SDSSJ1153+4612 &   3  &   3  &   0  &   3  &  1.7 &  1.0 &    -- \\
SDSSJ1205+4910 &   3  &   3  &   3  &   3  &  1.9 &  1.9 & F555W \\
SDSSJ1213+6708 &   1  &   1  &   0  &   3  &  2.3 &  1.9 &    -- \\
SDSSJ1218+0830 &   3  &   3  &   0  &   3  &  2.9 &  0.7 &    -- \\
SDSSJ1250+0523 &   3  &   3  &   0  &   3  &  1.9 &  1.9 &    -- \\
SDSSJ1402+6321 &   3  &   3  &   0  &   3  &  1.6 &  0.4 & F555W \\
SDSSJ1416+5136 &   0  &   3  &   0  &   3  &  0   &  0   &    -- \\
SDSSJ1420+6019 &   3  &   3  &   0  &   3  &  2.1 &  0.1 &    -- \\
SDSSJ1430+4105 &   3  &   3  &   1  &   3  &  1.9 &  1.7 &    -- \\
SDSSJ1432+6317 &   3  &   2  &   0  &   3  &  2.9 &  2.1 &    -- \\
SDSSJ1451-0239 &   0  &   3  &   2  &   3  &  0   &  0   & F555W \\
SDSSJ1525+3327 &   2  &   2  &   0  &   2  &  2.0 &  1.9 &    -- \\
SDSSJ1627-0053 &   3  &   3  &   2  &   3  &  2.9 &  2.1 & F555W \\
SDSSJ1630+4520 &   3  &   3  &   1  &   3  &  1.9 &  0.2 & F555W \\
SDSSJ2238-0754 &   3  &   3  &   1  &   3  &  2.2 &  0.0 & F555W \\
SDSSJ2300+0022 &   3  &   3  &   0  &   3  &  1.8 &  0.0 & F555W \\
SDSSJ2303+1422 &   3  &   3  &   0  &   3  &  1.8 &  0.0 & F555W \\
SDSSJ2341+0000 &   3  &   3  &   0  &   3  &  2.2 &  1.7 &    -- \\
\hline
\multicolumn{8}{l}{\scriptsize\raggedright Note: $H_1 = 0$ indicates that the 
system was not inspected;}\\
\multicolumn{8}{l}{\scriptsize\raggedright these objects have $\hro < 1.5$}\\
\end{tabular}
\end{center}
\end{table*}

\begin{figure*}
\centering\epsfig{file=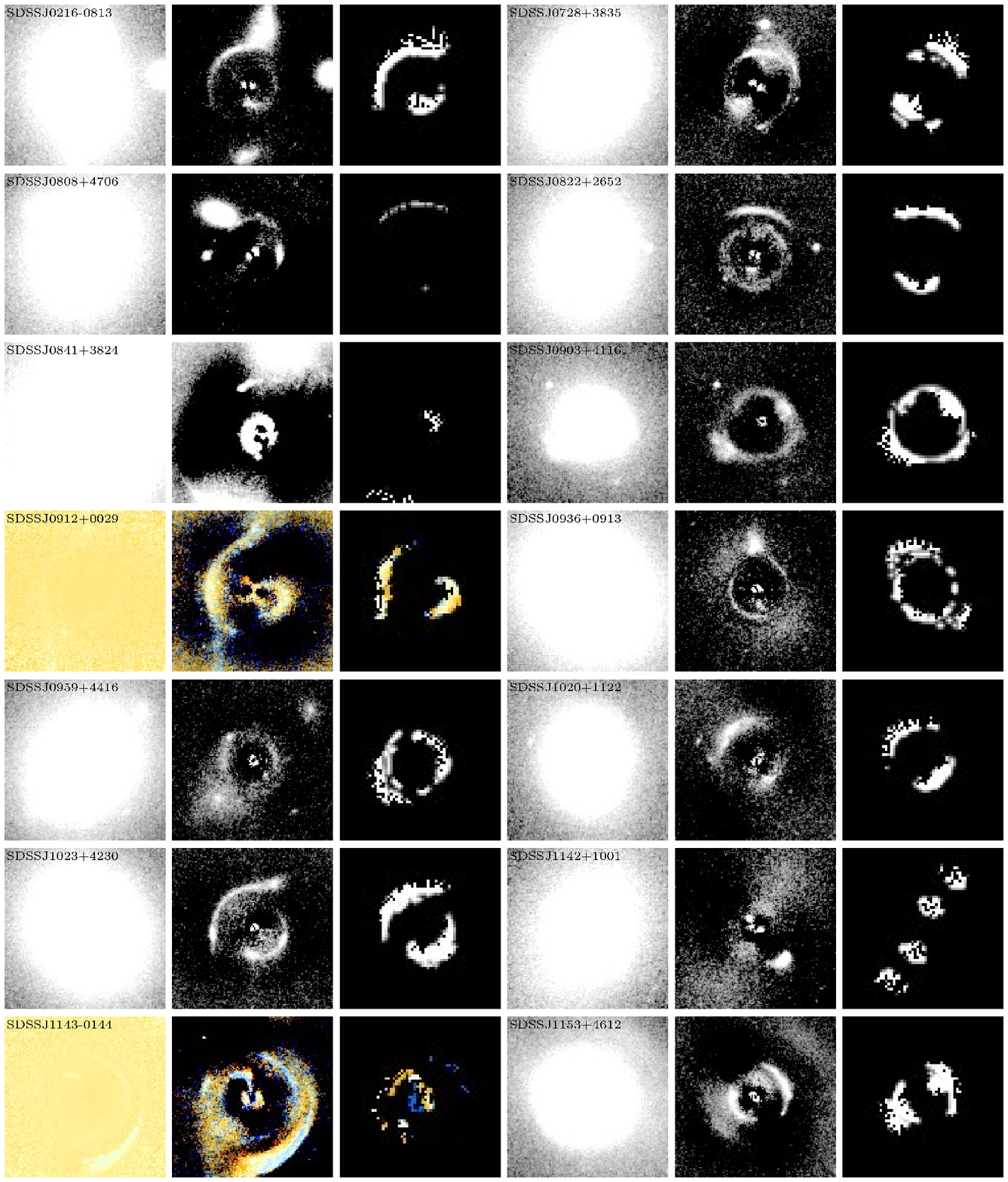,width=0.9\linewidth}
\caption{Robot output for SLACS main lenses with $1.5 < \hro < 2.5$, including 
galaxy cutouts, subtraction residuals, and predicted images from the 
robot lens models.\label{fig:mainlensesB}}
\end{figure*}

\begin{figure*}
\centering\epsfig{file=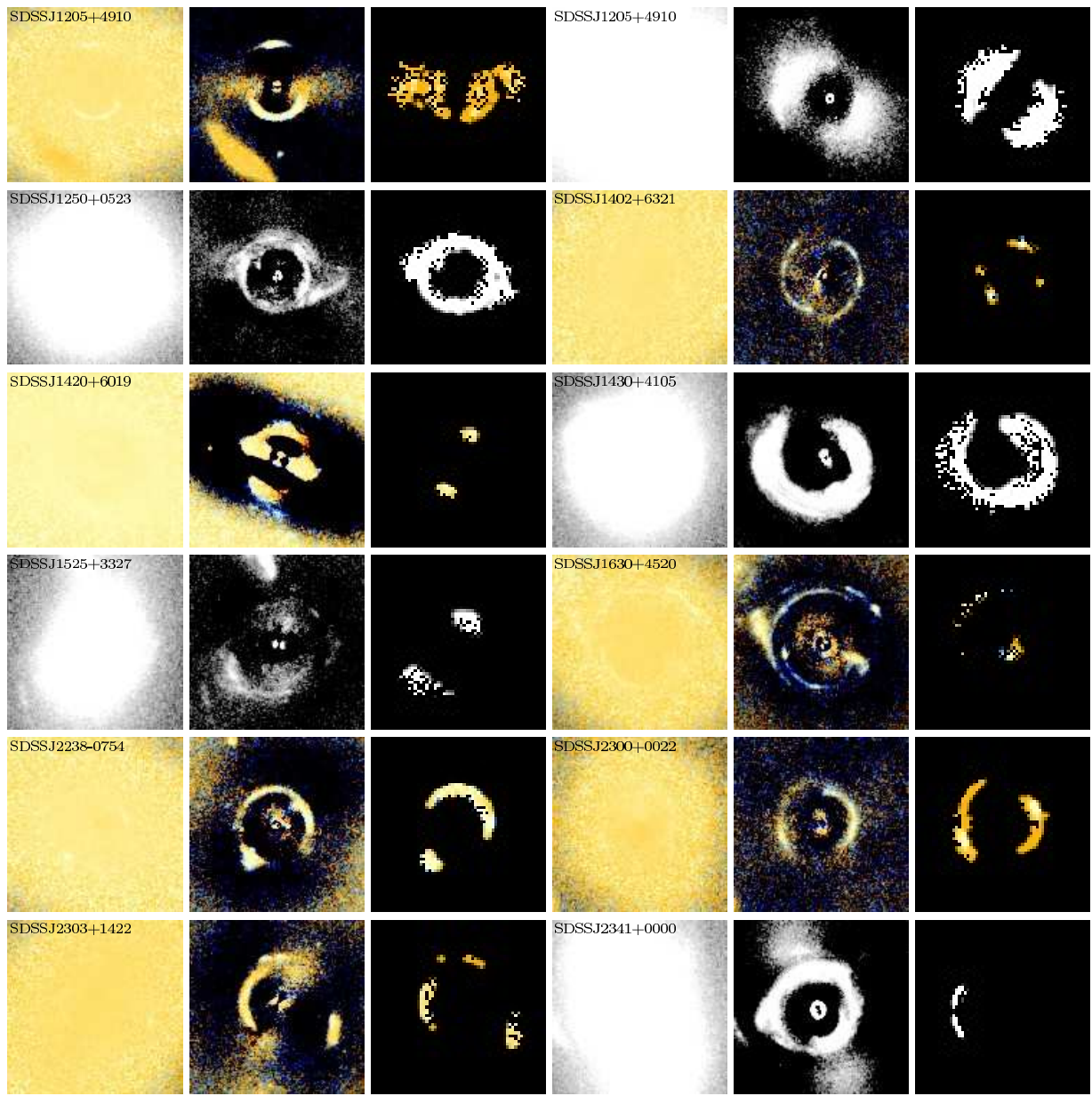,width=0.9\linewidth}
\caption{Continuation of Figure~\ref{fig:mainlensesB}. \label{fig:mainlensesBb}}
\end{figure*}

% ----------------------------------------------------------------------------

\bibliographystyle{apj}

\end{document}